______________________________________________________________________________
\documentstyle[equations,12pt]{article}
\renewcommand{\theequation}{\arabic{section}.\arabic{equation}}

\begin{document}

\begin{flushright}
IPNO/TH 96-01
\end{flushright}
\vspace{1 cm}
\begin{center}
{\large {\bf The relativistic two-body potentials of constraint theory\\
from summation of Feynman diagrams}}
\vspace{1.5 cm}

H. Jallouli\footnote{e-mail: jallouli@ipncls.in2p3.fr} 
and H. Sazdjian\footnote{e-mail: sazdjian@ipncls.in2p3.fr}\\
\renewcommand{\thefootnote}{\fnsymbol{footnote}}
{\it Division de Physique Th\'eorique\footnote{Unit\'e de Recherche
des Universit\'es Paris 11 et Paris 6 associ\'ee au CNRS.},
Institut de Physique Nucl\'eaire,\\
Universit\'e Paris XI,\\
F-91406 Orsay Cedex, France}\\
\end{center}
\newpage

\begin{center}
{\large Abstract}
\end{center}
The relativistic two-body potentials of constraint theory for systems
composed of two spin-0 or two spin-$\frac{1}{2}$ particles are calculated,
in perturbation theory, by means of the Lippmann$-$Schwinger type
equation that relates them to the scattering amplitude. The cases of
scalar and vector interactions with massless photons are considered. 
The two-photon exchange contributions, calculated with covariant 
propagators, are globally free of spurious infra-red singularities 
and produce at
leading order $O(\alpha ^4)$ effects that can be represented in
three-dimensional $x$-space by local potentials proportional to
$(\alpha /r)^2$. An approximation scheme, that adapts the eikonal
approximation to the bound state problem, is deviced and applied to the
evaluation of leading terms of higher order diagrams. Leading 
contributions of $n$-photon exchange diagrams produce terms proportional 
to $(\alpha /r)^n$. The series of leading contributions are summed.
The resulting potentials are functions, in the c.m. frame, of $r$
and of the total energy. Their forms are compatible with Todorov's
minimal substitution rules proposed in the quasipotential approach.
\par

\newpage


\section{Introduction}

The three-dimensional reduction of the Bethe$-$Salpeter equation 
\cite{sb,gml,nnns} was achieved in the past with the techniques of the
quasipotential approach \cite{ltlttk,bs,g,pl,f,fh,t1,lcl,mw}. The
method consisted in iterating the Bethe$-$Salpeter equation around 
a three-dimensional hypersurface in relative momentum space; the
result is a three-dimensional wave equation for the relative motion,
where the potential is related to the scattering amplitude by means
of a Lippmann-Schwinger type equation. This wave equation is not
unique in form and depends among others on the three-dimensional 
hypersurface chosen for the projection operation, reflecting the way
of eliminating the relative energy and relative time variables from
the initial four-dimensional equation. A variant of this approach
consisted in iterating the Bethe$-$Salpeter kernel around a simpler
three-dimensional one \cite{by,brbr}.
\par
Quantitative applications of these three-dimensional methods concerned
mainly QED and the positronium system. It turns out that the Coulomb
gauge is the most convenient gauge to achieve the perturbative
calculations \cite{lcl,by,brbr,m,n}. Covariant gauges \cite{l}, or
covariant propagators for scalar exchanges \cite{bcm}, have the
tendancy to produce spurious infra-red singularities that are only
cancelled by contributions of higher order diagrams. This feature
considerably reduces the domain of practical applicability of the
Bethe$-$Salpeter equation in covariant perturbative calculations.
Actually, this flaw can be traced back to the noncovariant nature of
the three-dimensional equation around which iteration is accomplished.
Also, the instantaneous approximation \cite{s}, applied to covariant
propagators, does not produce the correct $O(\alpha ^4)$ terms in
perturbation theory, neither the correct infinite mass limit, unless
an infinite number of higher order diagrams are taken into account
\cite{b}.
\par
In this respect, the variant of the quasipotential approach, developed
by Todorov \cite{t1}, where, to some extent, potentials may appear in
local form in $x$-space with appropriate c.m. total energy dependences,
provides a manifestly covariant framework that is free of the
abovementioned diseases. In particular, it has been shown \cite{rta},
for one spin-$\frac {1}{2}$ and one spin-0 particle systems, that in
the two-photon exchange diagrams, calculated in the Feynman gauge,
the spurious infra-red singularities cancel out and correct results are
obtained for the $O(\alpha ^4)$ and $O(\alpha ^5 \ln \alpha ^{-1})$
effects. This approach could not, however, reach sufficient generality, 
because of the lack in it of covariant wave equations for 
two-spin-$\frac {1}{2}$ particle systems.
\par
It is the manifestly covariant formalism of constraint theory
\cite{ll}, applied to two-particle systems \cite{dv,t2,k},
that provided the general framework for setting up covariant wave
equations with the correct number of degrees of freedom. 
Two-spin-$\frac {1}{2}$ fermion systems are described there by two
compatible Dirac type equations \cite{ls,cva1,s1}. In local approximations
of the potentials, these equations, without loss of their covariance
property, can be reduced to a single Pauli-Schr\"odinger type 
equation \cite{ms1} which displays their practical applicability to
spectroscopic problems. The connection of these equations with the
Bethe$-$Salpeter equation is obtained with standard iteration 
techniques \cite {s2,bb1}. (For a comparison of various 
three-dimensional equations see Ref. \cite{bb2}.)
\par
The purpose of the present paper is to investigate the perturbation
theory properties of the potentials of constraint theory wave equations.
We shall not consider here radiative corrections, which actually are
calculable independently from any three-dimensional formalism, but
rather shall analyze the structure of multiphoton-exchange diagrams.
\par
Our analysis is accomplished in three steps. Firstly, we study the
structure of two-photon exchange diagrams and show that the corresponding
potential is free of spurious infra-red logarithmic singularities
and yields the correct $O(\alpha ^4)$ effects; it can be represented,
in three-dimensional $x$-space, by means of a local fuction,
proportional to $(\alpha /r)^2$. Secondly, we device an approximation
method for the evaluation of the above diagrams that globally reproduces 
the same result for their leading effect. This method is
a variant of the eikonal approximation \cite{cw,lesu,aibizj}, adapted
to the bound state problem.
\par
Thirdly, we apply the approximation method to the higher order 
diagrams. We show that the $n$-photon exchange diagrams yield for the
corresponding potential, as a dominant contribution, an 
$O(\alpha ^{2n})$ effect, represented by a term proportional to
$(\alpha /r)^n$. We then sum the series of dominant contributions to 
obtain the potential in compact form, as a function of $(\alpha /r)$
and of the c.m. total energy.
\par
The expression we find for the timelike component of the electromagnetic
potential can naturally be completed to include the spacelike components
as well. It then represents the fermionic generalization of the 
potential proposed by Todorov in his quasipotential approach for 
two-spin-0 particle systems, on the basis of minimal substitution rules 
\cite{t1}.
The same calculations, applied to scalar photons, yield a potential
that is equivalent, in the static approximation of its higher order
terms, to the form proposed by Crater and Van Alstine \cite{cva1},
who extended Todorov's substitution rules to scalar interactions.
\par
The above calculations result in sizable effects mainly in strong
coupling problems. One such problem is provided by the evaluation
of the short-distance potential in QCD, which is very similar to
the QED potential, with values of the coupling constant of the 
order of 0.5; furthermore, part of the radiative
corrections could be taken into account by the replacement of the
coupling constant by the effective coupling constant, obtained from
renormalization group analysis.
\par
Another domain of application of these potentials is provided by
QED in its strong coupling regime, where lattice calculations signal
a chiral phase transition for values of $\alpha $ of the order of
0.3-0.4 \cite{br,go}. Other qualitative results obtained from
the Bethe$-$Salpeter equation with one-photon or one-gluon
exchanges (such as spontaneous breakdown of chiral symmetry) might be
tested with these potentials and their stability in the presence of
multiphoton or multigluon exchanges checked. Also, these potentials
could naturally be generalized for the treatment of other
interactions, such as those corresponding to the exchanges of massive
particles or to confining interactions, by the replacement there of
the Coulomb potential by another appropriate potential.
\par
The paper is organized as follows. In Sec. 2 the wave equations of 
constraint theory are introduced. In Sec. 3, their connection with
the Bethe$-$Salpeter equation and the relationship of the potential
with the scattering amplitude are established. In Sec. 4, the 
perturbative calculational method of the potential is developed.
In Sec. 5, the potential corresponding to the exchanges of two
scalar photons for two-spin-0 particle systems is calculated to order
$\alpha ^4$. In Sec. 6, similar calculations are done for scalar and 
vector photons for a fermion-antifermion system. In Sec. 7, the 
approximation method leading to the results of Secs. 5 and 6 is
deviced and some of its general consequences about cancellation of
diagrams are shown. In Secs. 8 and 9, the dominant contributions coming 
from multiphoton exchanges are calculated and summed for two-boson and 
fermion-antifermion systems, respectively. In Sec. 10, the Bethe$-$Salpeter 
wave function is reconstituted in terms of the constraint theory wave 
function in the framework of the present approximations. Conclusion 
follows in Sec. 11. Details of the calculations of two-photon exchange 
diagrams are presented in Appendices A and B. A summary of the 
calculation of the vector potential was presented by the present 
authors in Ref. \cite{js}.
\par

\newpage                                                   

\section{Two-body wave equations of constraint theory}
\setcounter{equation}{0}
                                               
A system of two-spin-0 particles, with masses $m_1$ and $m_2$, 
respectively, is described by means of two wave equations that are
generalizations of the Klein$-$Gordon equation \cite{dv,t2,k,cva1,s1}:
\subequations
\begin{eqnarray}
\label{2e1a}
(p_1^2 - m_1^2 - V)\ \Psi (x_1,x_2) &=& 0\ ,\\
\label{2e1b}
(p_2^2 - m_2^2 - V)\ \Psi (x_1,x_2) &=& 0\ ,  
\end{eqnarray}
\endsubequations
where $V$ is a Poincar\'e invariant function or operator of the 
variables of the system. It is possible to choose different potentials in 
the two equations, but by means of canonical (or wave function) 
transformations one can bring the corresponding equations to the above
configuration \cite{s3}.
\par
We use the following definitions for the total and relative variables:
\begin{eqnarray} \label{2e2}
P &=& p_1 + p_2\ ,\ \ \ p\ =\ \frac {1}{2}(p_1-p_2)\ ,\ \ \ 
M\ =\ m_1 + m_2\ ,\nonumber \\
X &=& \frac {1}{2}(x_1+x_2)\ ,\ \ \ x\ =\ x_1-x_2\ .
\end{eqnarray}
For states that are eigenstates of the total momentum $P$ we define
transverse and longitudinal components of four-vectors with respect
to $P$:
\begin{eqnarray}  \label{2e3}
q_{\mu}^T &=& q_{\mu} - \frac {q.P}{P^2} P_{\mu}\ ,\ \ \ 
q_{\mu}^L\ =\ (q.\hat P) \hat P_{\mu}\ ,\ \ \ \hat P_{\mu}
\ =\ \frac {P_{\mu}}{\sqrt {P^2}}\ ,\ \ \ q_L\ =\ q.\hat P\ ,
\nonumber \\
P_L &=& \sqrt {P^2}\ ,\ \ \ \ q_L \big 
\vert _{c.m.} =\ q_0\ ,\ \ \ q^{T2}\big \vert _{c.m.} =\ 
-{\bf q^2}\ ,\ \ \ r\ =\ \sqrt {-x^{T2}}\ .
\end{eqnarray}
This decomposition is manifestly covariant. In the c.m. frame the
transverse components reduce to the three space components, while the
longitudinal component reduces to the time component of the 
corresponding four-vector.
\par
The compatibility condition of the two wave equations 
(\ref{2e1a})-(\ref{2e1b}) imposes the following constraints on the
wave function and the potential:
\begin{eqnarray}
\label{2e4}
& & [\ (p_1^2-p_2^2) - (m_1^2-m_2^2)\ ]\ \Psi \ =\ 0\ ,\\   
\label{2e5}
& &\big [\ p_1^2-p_2^2 , V\ \big ]\ \Psi \ =\ 0\ .
\end{eqnarray}
\par
The solution of Eq. (\ref{2e4}) is (for eigenfunctions of the total
momentum $P$):
\begin{equation} \label{2e6}
\Psi \ =\ e^{\displaystyle {-iP.X}}e^{\displaystyle 
{-i(m_1^2-m_2^2)x_L/(2P_L)}}\ \psi (x^T)\ ,
\end{equation}
while Eq. (\ref{2e5}) tells us that $V$ does not depend on the 
relative longitudinal coordinate $x_L$:
\begin{equation} \label{2e7}
V\ =\ V(x^T,P_L,p^T)\ .
\end{equation}
Thus, Eqs.
(\ref{2e4})-(\ref{2e5}) allow us to eliminate the longitudinal relative
coordinate and momentum. Equation (\ref{2e4}) can be rewritten in a
slightly different form by defining the constraint $C(p)$:
\begin{equation} \label{2e8}
C(p)\ \equiv \ 2P_Lp_L - (m_1^2-m_2^2)\ \approx \ 0\ .
\end{equation}
It yields for the individual logitudinal momenta the expressions:
\begin{equation} \label{2e9}
p_{1L}\ =\ \frac {P_L}{2} + \frac {(m_1^2-m_2^2)}{2P_L}\ ,\ \ \ \ 
p_{2L}\ =\ \frac {P_L}{2} - \frac {(m_1^2-m_2^2)}{2P_L}\ .
\end{equation}
\par
The eigenvalue equation satisfied by the internal wave function
$\psi $ is:
\begin{equation} \label{2e10}
\big [\ \frac{P^2}{4} - \frac {1}{2} (m_1^2+m_2^2) + \frac
{(m_1^2-m_2^2)^2}{4P^2} + p^{T2}\ \big ]\ \psi (x^T)\ =\ 0\ ,
\end{equation}
which is a three-dimensional Klein$-$Gordon$-$Schr\"odinger type
equation. Also notice that when constraint $C$ (\ref{2e8}) is
utilized, the two individual Klein$-$Gordon operators become equal:
\begin{equation} \label{2e11}
H_0\ \equiv \ (p_1^2-m_1^2)\big \vert _C \ =\ (p_2^2-m_2^2)\big
\vert _C \ =\ \frac {P^2}{4} - \frac {1}{2} (m_1^2+m_2^2) +
\frac {(m_1^2-m_2^2)^2}{4P^2} + p^{T2}\ .
\end{equation}
\par 
In general the potential $V$ [Eq. (\ref{2e7})] is a Poincar\'e 
invariant integral operator in $x^T$, but in simplified cases it can
be represented as a function of $x^T$ with a general dependence on
$P_L$, and an eventual quadratic dependence on $p^T$.
\par
The scalar product for the wave functions satisfying Eqs. 
(\ref{2e1a})-(\ref{2e1b}) can be obtained either from the 
construction of tensor currents satisfying two independent
conservation laws \cite{s1} with respect to $x_1$ and $x_2$,
respectively, or directly from the Green's function of Eqs.
(\ref{2e1a})-(\ref{2e1b}) or (\ref{2e10}) \cite{f,lcl}.
The result for the norm is (in the c.m. frame):
\begin{equation} \label{2e12}
\int d^3 {\bf x}\ \psi ^* 4P^2 \big [\ \frac {\partial }{\partial P^2}
(H_0-V)\ \big ] \psi \ =\ 2P_0\ ,
\end{equation}
where $H_0$ is the Klein$-$Gordon operator (\ref{2e11}).
\par
For a system of two-spin-$\frac {1}{2}$ particles, composed of one
fermion with mass $m_1$ and one antifermion with mass $m_2$, the 
constraint theory wave equations can be written in the form \cite{s1}:
\subequations
\begin{eqnarray}
\label{2e13a}
(\gamma _1.p_1-m_1)\widetilde \Psi &=& (-\gamma _2.p_2+m_2)
\widetilde V \widetilde \Psi\ ,\\
\label{2e13b}
(-\gamma _2.p_2-m_2)\widetilde \Psi &=& (\gamma _1.p_1+m_1)
\widetilde V \widetilde \Psi\ ,
\end{eqnarray}
\endsubequations              
where $\widetilde V$ is a Poincar\'e invariant potential. Here, 
$\widetilde \Psi$ is a sixteen-component spinor wave function of 
rank two and is represented as a $4\times 4$ matrix:
\begin{equation} \label{2e14}
\widetilde \Psi \ =\ \widetilde \Psi _{\alpha _1 \alpha _2}
(x_1,x_2)\ \ \ \ \ (\alpha _1,\alpha _2 =1,\cdots ,4)\ ,
\end{equation}
where $\alpha _1\ (\alpha _2)$ is the spinor index of particle 1 (2).
$\gamma _1$ is the Dirac matrix $\gamma $ acting in the subspace of the 
spinor of particle 1 (index $\alpha _1$); it acts on $\widetilde
\Psi $ from the left. $\gamma _2$ is the Dirac matrix acting in the
subspace of the spinor of particle 2 (index $\alpha _2$); it acts on 
$\widetilde \Psi $ from the right; this is also the case of products
of $\gamma _2$ matrices, which act on $\widetilde \Psi$ from the right
in the reverse order:
\begin{eqnarray} \label{2e15}
& &\gamma _{1\mu} \widetilde \Psi \ \equiv \ (\gamma _{\mu})_
{\alpha _1 \beta _1} \widetilde \Psi _{\beta _1 \alpha _2}\ ,\ \ \ \
\gamma _{2\mu} \widetilde \Psi \ \equiv \ \widetilde \Psi _
{\alpha _1 \beta _2} (\gamma _{\mu})_{\beta _2 \alpha _2}\ , 
\nonumber \\
& &\gamma _{2\mu} \gamma _{2\nu} \widetilde \Psi \ \equiv \
\widetilde \Psi _{\alpha _1 \beta _2} (\gamma _{\nu} \gamma _{\mu})_
{\beta _2 \alpha _2}\ ,\ \ \ \ \sigma _{a\alpha \beta}\ \equiv \
\frac {1}{2i} \big [ \gamma _{a\alpha} , \gamma _{a\beta}
\big ]\ \ \ (a=1,2)\ .
\end{eqnarray}
\par  
The compatibility condition of Eqs. (\ref{2e13a})-(\ref{2e13b}) yields 
the same constraints as in Eqs. (\ref{2e4}) and (\ref{2e5}) (with 
$\Psi$ and $V$ replaced by $\widetilde \Psi$ and $\widetilde V$,
respectively). In particular, one has:
\begin{equation} \label{2e16}
\widetilde V\ =\ \widetilde V(x^T,P_L,p^T,\gamma _1,\gamma _2)\ .
\end{equation}
\par
Using the operator $H_0$ [Eq. (\ref{2e11})], Eqs. 
(\ref{2e13a})-(\ref{2e13b}) can also be rewritten as a single 
equation, provided the constraint (\ref{2e8}) is used:
\begin{equation} \label{2e17}
\bigg [\ \frac {1}{i^2} (\gamma _1.p_1-m_1) (-\gamma _2.p_2-m_2)
H_0^{-1}\ \bigg ]_{C(p)} \widetilde \Psi \ \equiv \ \bigg [ S_1(p_1)
S_2(-p_2) H_0 \bigg ]_{C(p)}^{-1}\widetilde \Psi \ =\
-\widetilde V \widetilde \Psi \ ,
\end{equation}
where $S_1$ and $S_2$ are the propagators of fermions 1 and 2,
respectively.
\par
The norm for the internal wave function $\widetilde \psi$ takes the
form (in the c.m. frame):
\begin{equation} \label{2e18}
\int d^3{\bf x}\ Tr \bigg \{ \widetilde \psi ^{\dagger}\big [ 1-
\widetilde V^{\dagger}\widetilde V + 4\gamma _{10}\gamma _{20} P_0^2
\frac {\partial \widetilde V}{\partial P^2}\big ] \widetilde \psi
\bigg \}\ =\ 2P_0\ ,
\end{equation}
with $\widetilde V$ satisfying the hermiticity condition
\begin{equation} \label{2e19}
\widetilde V^{\dagger}\ =\ \gamma _{10}\gamma _{20} \widetilde V
\gamma _{10}\gamma _{20}\ .
\end{equation}
\par
Equation (\ref{2e18}) shows that, for energy independent potentials
(in the c.m. frame) the norm of $\widetilde \psi$ is not positive 
definite. In order to ensure positivity, it is sufficient that the 
potential $\widetilde V$ satisfy the inequality
\begin{equation} \label{2e20}
\frac {1}{4} Tr (\widetilde V^{\dagger}\widetilde V) \ < \ 1\ .
\end{equation}
In this case one is allowed to make the wave function transformation
\begin{equation} \label{2e21}
\widetilde \Psi\ =\ \big [1-\widetilde V^{\dagger}\widetilde V\big ]^
{-1/2} \Psi
\end{equation}
and to reach a representation where the norm for c.m. energy independent
potentials is the free norm.
\par
In this respect, the parametrization suggested by Crater and Van Alstine
\cite{cva2}, for local potentials that commute with $\gamma _{1L}
\gamma _{2L}$ (hence $\widetilde V^{\dagger}=\widetilde V$),
\begin{equation} \label{2e22}
\widetilde V\ =\ \tanh V\ ,
\end{equation}
satisfies condition (\ref{2e20}) and allows one to bring the equations
satisfied by $\Psi$ [Eq. (\ref{2e21})] into forms analogous to the
Dirac equation, where each particle appears as placed in the external 
potential created by the other particle, the latter potential having
the same tensor nature as potential $V$ of Eq. (\ref{2e22}).
\par
We shall henceforth adopt parametrization (\ref{2e22}) for potential 
$\widetilde V$ and, according to Eq. (\ref{2e21}), shall introduce
the wave function transformation
\begin{equation} \label{2e23}
\widetilde \Psi\ =\ (\cosh V)\Psi\ .
\end{equation}
[For potentials that do not commute with $\gamma _{1L}\gamma _{2L}$,
the generalizations of Eqs. (\ref{2e22}) and (\ref{2e23}) and of the
corresponding norm are presented in Ref. \cite{ms2}.] The norm of the
internal part of the new wave function $\Psi$ then becomes (in the 
c.m. frame):
\begin{equation} \label{2e24}
\int d^3{\bf x}\ Tr \bigg \{ \psi ^{\dagger}\big [ 1+4\gamma _{10}
\gamma _{20} P_0^2 \frac {\partial V}{\partial P^2}\big ]\psi
\bigg \}\ =\ 2P_0\ .
\end{equation}
[The relationship between $\Psi$ and $\psi$ is the same as in Eq.
(\ref{2e6}).]
\par
In order to obtain the final eigenvalue equation of $\psi$, one has to
decompose the latter along $2\times 2$ components. We choose, for this
decomposition, the basis of the matrices $1$, $\gamma _L$, $\gamma _5$
and $\gamma _L\gamma _5$:
\begin{equation} \label{2e25}
\psi\ =\ \psi _1\ +\ \gamma _L\psi _2\ +\ \gamma _5\psi _3\ +\ \gamma _L
\gamma _5 \psi _4\ .
\end{equation}
\par
When potential $V$ is a function of $x^T$ (i.e., not an integral operator)
it is possible, for some general classes of potential, to eliminate
through the wave equations (\ref{2e13a})-(\ref{2e13b}) three of these
components in terms of a fourth one. We choose $\psi _3$ for the
remaining component, which is also a surviving component in the 
nonrelativistic limit. For a general combination of scalar, pseudoscalar
and vector potentials, the final eigenvalue equation is of the 
Pauli$-$Schr\"odinger type, the radial momentum operator appearing in it
through the Laplace operator. For this class of potential, $V$ is decomposed    
along the $\gamma $ matrices as:
\begin{equation} \label{2e26}
V\ =\ V_1 + \gamma _{15}\gamma _{25}V_3 + \gamma _1^{\mu}\gamma _2^
{\nu} (g_{\mu\nu}^{LL}V_2 + g_{\mu\nu}^{TT}U_4 + \frac {x_{\mu}^T
x_{\nu}^T}{x^{T2}}T_4)\ ,
\end{equation}
where $V_1$ is the scalar, $V_3$ the pseudoscalar and $V_2$ the timelike
vector potential; $U_4$ and $T_4$ are the spacelike vector potentials. 
[The numbering of the indices and the notations are related to the 
particular projectors to which these potentials are attached \cite{ms1}.]
\par
After making a change of function of the type
\begin{equation} \label{2e28}
\psi _3\ =\ A\phi _3\ ,   
\end{equation}
the expression of $A$ being given in Ref. \cite{ms1}, one finds 
the following Pauli$-$Schr\"odinger type equation (in the
c.m. frame):
\begin{eqnarray} \label{2e29}
\bigg \{e^{{\displaystyle 4(U_4+T_4)}}\ \big [\ \frac {P^2}{4}
e^{{\displaystyle 4V_2}}&-& \frac {M^2}{4}e^{{\displaystyle 4V_1}}
- \frac {(m_1^2-m_2^2)^2}{4M^2}e^{{\displaystyle -4V_1}} 
+ \frac {(m_1^2-m_2^2)^2}{4P^2}e^{{\displaystyle -4V_2}}\ \big ]
\nonumber \\
&-& {\bf p^2}\ -\ {\bf L^2}\frac {1}{{\bf x^2}} (e^{{\displaystyle 4T_4}}
-1)\ +\ \cdots \bigg \}\ \phi_3\ =\ 0\ , 
\end{eqnarray}
where the dots stand for the various spin dependent or 
$\hbar ^2$-dependent terms and ${\bf L}$ is the orbital angular momentum 
operator. [The complete form of the equation can be found in Ref.
\cite{ms1}.] We recognize that the scalar potential $V_1$ acts as a 
modification of the total mass $M$ of the fermions through the change 
$M\rightarrow Me^{{\displaystyle 2V_1}}$ while $(m_1^2-m_2^2)$ is kept 
fixed. The timelike vector potential $V_2$ acts as a modification of the 
c.m. total energy $P_L$ through the change $P_L\rightarrow P_L
e^{{\displaystyle 2V_2}}$, while $(p_{1L}^2-p_{2L}^2)=(m_1^2-m_2^2)$ is 
kept fixed. The nonmodification of $(m_1^2-m_2^2)$ is simply due to the 
constraint (\ref{2e8}). The spacelike potential $U_4$ changes the orbital 
angular momentum operator from ${\bf L}$ to 
${\bf L}e^{{\displaystyle -2U_4}}$ and the combination $U_4+T_4$ of the
spacelike potentials changes the radial momentum operator from $p_r$ to 
$p_r e^{{\displaystyle -2(U_4+T_4)}}$ (in the classical limit). The 
pseudoscalar potential appears only in spin dependent and 
$\hbar ^2$-dependent terms.
All these effects are what one expects from an external field
interpretation of the potentials in the Dirac and Klein$-$Gordon 
equations and therefore provide additional justification for the
parametrization (\ref{2e22}) and the decomposition of $V$ along the
form (\ref{2e26}).
\par
Wave equations can also be written down for a system made of one 
spin-$\frac {1}{2}$ and one spin-0 particles \cite{s1}. We shall not,
however, consider such a system in the present work.
\par

\newpage
   
\section{Connection with the Bethe$-$Salpeter equation \protect \\
and the scattering amplitude}
\setcounter{equation}{0}

In order to find the connection of the wave equations 
(\ref{2e1a})-(\ref{2e1b}) and
(\ref{2e13a})-(\ref{2e13b}) with the Bethe$-$Sal\-peter equation, it is
natural to project, with an appropriate weight factor, 
four-dimensional quantities such as Green's functions, scattering
amplitudes and wave functions on the constraint hypersurface (\ref{2e8})
and to iterate the corresponding integral equations around that 
hypersurface. There is an arbitrariness in the choice of the weight
factor, but it is limited by several additional conditions, such as
correct one-body limit, absence of spurious singularities, correct
$O(\alpha ^4)$ effects, interchange symmetry between particles 1 and 2, 
hermiticity of the potential, etc.. We shall first present the general 
method of approach and then shall discuss the question of the choice of 
the weight factor.
\par
Let $G(P,p,p')$ be the four-point Green's function (for bosons or
fermions) with total momentum $P$ and relative momenta $p'$ and $p$
for the ingoing and outgoing particles, respectively [Eqs. (\ref{2e2})],
and from which the total four-momentum conservation factor
$(2\pi)^4 \delta ^4 (P-P')$ has been amputated. It satisfies the
integral equation:
\begin{equation} \label{3e1}
G\ =\ G_0 + G_0KG\ ,
\end{equation}
where $G_0$ is the free two-particle propagator,
\begin{equation} \label{3e2}
G_0\ =\ G_1 G_2
\end{equation}
[$G_1$ and $G_2$ are the propagators of particles 1 and 2,
respectively], and $K$ is the Bethe-Salpeter kernel \cite{sb,gml,nnns}.
The off-mass shell scattering amplitude $T$ (from which the total 
four-momentum conservation factor has been amputated) is related to $K$ 
by the relation:
\begin{equation} \label{3e3}
T\ =\ K + KG_0T\ .
\end{equation}
\par
In the vicinity of a bound state pole, $G$ behaves as 
\begin{equation} \label{3e4}
\lim _{s\rightarrow s_B} G(P,p,p')\ =\ i\eta \frac
{\phi (P,p) \overline \phi (P,p')}{s-s_B+i\epsilon}\ ,\ \ \ \ s\ =\ P^2\ ,
\ \ \ \eta = \left \{ \begin{array}{ll} +1 & {\rm for\  bosons,}\\
-1 & {\rm for\ fermions,} \end{array} \right. 
\end{equation}                                 
where $\phi$ is the internal part of the Bethe$-$Salpeter wave function
and $\overline \phi$ is defined from hermitian conjugation and
antichronological product.
\par
We next define the left-projected Green's function on the hypersurface
(\ref{2e8}):
\begin{equation} \label{3e5}
\widetilde G(p,p')\ =\ -2i\pi\delta (C(p)) G_{0C}^{-1} G(p.p')\ ,
\end{equation}
where $G_{0C}^{-1}$ is the weight factor to be determined below. 
[When not necessary, total four-momentum will not be explicitly written
in the arguments.] Notice that $\widetilde G$ is still
four-dimensional with respect to its right argument.
\par
It is now possible to iterate $G$ around $\widetilde G$ and obtain
the integral equation for $\widetilde G$. We shall use the following
definitions and abbreviated notations:
\begin{eqnarray} \label{3e6}
\delta _C &=& \delta (C(p))\ , \ \ \ \ \ \ \
g\ =\ G_0 + 2i\pi \delta _C G_{0C}^{-1} G_0\ ,\nonumber \\
\widetilde g &=& +2i\pi \delta _C G_{0C}^{-1} G_0\ , \ \ \ \ \ \ \ 
\widetilde g_0 = G_{0C}^{-1} G_0 \big \vert _{C(p)}\ .       
\end{eqnarray}
\par
Using in the right-hand side of Eq. (\ref{3e5}) the integral equation
for $G$, one obtains for the difference $G-\widetilde G$:
\begin{equation} \label{3e7}
G-\widetilde G\ =\ g(1+KG)\ ,
\end{equation}
which yields:
\begin{equation} \label{3e8}
G\ =\ (1-gK)^{-1} (\widetilde G + g)\ .
\end{equation}
One then successively uses Eqs. (\ref{3e1}) and (\ref{3e8}) in the
right-hand side of Eq. (\ref{3e5}) to obtain:
\begin{equation} \label{3e9}
\widetilde G\ =\ -\widetilde g \big [ (1-Kg)^{-1} + K(1-gK)^{-1} 
\widetilde G \big ]\ .
\end{equation}
The second term in the right-hand side of this equation can also be
expressed in terms of the scattering amplitude with the use of Eq.
(\ref{3e3}):
\begin{equation} \label{3e10}
\widetilde G\ =\ -\widetilde g\big [ (1-Kg)^{-1} + T(1-\widetilde gT)^
{-1}\widetilde G\big ]\ .
\end{equation}
\par
We define the constraint theory wave function $\psi$ with the same 
projection as in Eq. (\ref{3e5}):
\begin{equation} \label{3e11}
\psi =\ -2i\pi \delta _C G_{0C}^{-1} \phi\ ,
\end{equation}
which leads to the following behavior of $\widetilde G$ in the 
neighberhood of a bound state pole:
\begin{equation} \label{3e12}
\lim _{s\rightarrow s_B} \widetilde G(p,p')\ =\ i\eta \frac {\psi (p)
\overline \phi (p')}{s-s_B+i\epsilon}
\end{equation}
[$\eta$ is defined in Eq. (\ref{3e4})], which, in turn, yields the
wave equation satisfied by $\psi$:
\begin{equation} \label{3e12'}
\psi \ =\ -\widetilde g T(1-\widetilde g T)^{-1} \psi\ .
\end{equation}
\par
We shall also define the reduced wave function $\widetilde \psi$ and 
amplitude $\widetilde T$ as:
\subequations
\begin{eqnarray} 
\label{3e13b}
\psi &=& 2\pi 2P_L \delta _C \widetilde \psi\ ,\\
\label{3e13c}
\widetilde T(p,p') &=& \frac {i}{2P_L} T(p,p')\big \vert _{C(p),C(p')}\ .
\end{eqnarray}  
\endsubequations
\par
The wave equation satisfid by $\widetilde \psi$ takes the form:
\begin{equation} \label{3e14}
\widetilde g_0^{-1} \widetilde \psi \ =\ -\widetilde T (1-\widetilde g_0  
\widetilde T)^{-1} \widetilde \psi\ ,
\end{equation}
where now the integrations between $\widetilde T$, $\widetilde g_0$ and
$\widetilde \psi$ are three-dimensional and concern the transverse
variables [Eqs. (\ref{2e3})], after the constraint (\ref{2e8}) has
been used for the longitudinal momenta.
\par
It is also possible to reconstitute the Green's function $G$ and the
Bethe$-$Salpeter wave function $\phi$ from the knowledge of the 
constraint theory Green's function $\widetilde G$ and wave function
$\psi$. To this end, we consider Eq. (\ref{3e7}), in the right-hand
side of which we replace $G$ by its expression (\ref{3e8}), and use 
Eq. (\ref{3e9}) for $\widetilde G$; we find:
\begin{eqnarray} 
\label{3e15}
G &=& G_0 + G_0T(1-\widetilde gT)^{-1}(\widetilde G+g)\ ,\\
\label{3e16}
\phi &=& G_0T(1-\widetilde gT)^{-1} \psi \ .
\end{eqnarray}
Notice that in the left argument of the first $T$ in the above equations
the constraint (\ref{2e8}) is not used. We also dispose for $G$ of Eq. 
(\ref{3e8}), which yields for $\phi$ the relation  
\begin{equation} \label{3e17}
\phi\ =\ (1-gK)^{-1}\psi\ ,
\end{equation}
but these equations are more singular than Eqs. (\ref{3e15}) and
(\ref{3e16}) above, and should only be used inside integrals.
\par
It is natural at this stage to define the potential $\widetilde V$
(also denoted by $V$ for the bosonic case) by the Lippmann$-$Schwinger
type relation:
\begin{equation} \label{3e18}
\widetilde V\ =\ \widetilde T (1-\widetilde g_0\widetilde T)^{-1}\ .
\end{equation}
Equation (\ref{3e14}) then becomes:
\begin{equation} \label{3e19}
\widetilde g_0^{-1} \widetilde \psi\ =\ -\widetilde V \widetilde \psi\ .
\end{equation}
\par
This equation has the same structure as the constraint theory wave
equations (\ref{2e1a})-(\ref{2e1b}) and (\ref{2e17}). Comparison
of $\widetilde g_0^{-1}$ with the corresponding kinematic operators then 
fixes the expression of $G_{0C}^{-1}$ [Eqs. (\ref{3e5}) and (\ref{3e6})]:
\begin{equation} \label{3e20}
G_{0C}^{-1}\big \vert _{C(p)}=\ H_0\ .
\end{equation}
[$H_0$ is defined in Eq. (\ref{2e11}).]
\par
Let us now examine the hermiticity property of potential $\widetilde V$,
which should partly be guaranteed by the on-mass shell elastic unitarity
property of the scattering amplitude, through relation (\ref{3e18}).
After a little algebra \cite{t1}, one finds for the bosonic case the 
following condition on $\widetilde g_0$:
\begin{equation} \label{3e21}
\widetilde g_0 - \widetilde g_0^*\ =\ 2i\pi \delta (b_0^2(s) + 
p^{T2}) \theta (s-(m_1+m_2)^2)\ ,
\end{equation}
where we have defined
\begin{eqnarray} \label{3e22}
b_0^2(s) &=& \frac {1}{4} (s-(m_1+m_2)^2) (s-(m_1-m_2)^2)
\nonumber \\                                      
&=& \frac {s}{4} - \frac {1}{2} (m_1^2+m_2^2) + \frac {(m_1^2-m_2^2)^2}
{4s}\ .
\end{eqnarray}
The argument of the $\delta $-function above is nothing but the 
operator $H_0$ defined in Eq. (\ref{2e11}), which means that solution
(\ref{3e20}) for $G_{0C}^{-1}$ automatically satisfies the 
hermiticity-unitarity condition (\ref{3e21}) in the physical region
of the amplitude. [$G_{0C}$ should be defined with the prescription
$(H_0+i\epsilon)^{-1}$.]
\par
For the fermionic case, the hermiticity-unitarity condition reads:
\begin{eqnarray} \label{3e23}
\widetilde g_0 - \gamma _{1L}\gamma _{2L} \widetilde g_0^{\dagger}
\gamma _{1L}\gamma _{2L} &=& 2i\pi \delta (b_0^2(s)+p^{T2})
\theta (s-(m_1+m_2)^2) \nonumber \\
& & \times (\gamma _1.p_1+m_1) (-\gamma _2.p_2+m_2)\ ,
\end{eqnarray}
which is also satisfied by solution (\ref{3e20}).
\par 
Other solutions to Eqs. (\ref{3e21}) or (\ref{3e23}), different from
that given in Eq. (\ref{3e20}), can be constructed as well. In
principle, different choices of $\widetilde g_0$ should not affect
the physical predictions of the theory, but rather would amount to
organizing in different ways the resolution of the bound state 
problem and the summation of the perturbation series, putting
emphasis on specific approximations. In the peresent work, we stick
to the choice (\ref{3e20}), which has the two main advantages
of simplicity and of the fact that it reduces in $x$-space to a
second-order differential operator. It is only in this case that can
we device local approximation schemes for potential $\widetilde V$
and remain in a framework analogous to that of nonrelativistic
quantum mechanics. Choice (\ref{3e20}) has also the advantages of
hermiticity, interchange symmetry between particles 1 and 2 and 
correct one-body limit.
\par
We shall, however, slightly generalize solution (\ref{3e20}) by 
allowing in it the presence of a finite multiplicative 
renormalization constant. Such a generalization is necessary in the
off-mass shell formalism we are using to avoid the presence of
spurious $O(\alpha ^3)$-terms. It amounts to requiring that the 
$1/r$-terms of the potential result solely from the contribution
of the one-photon exchange diagram.
For this purpose we define
\begin{equation} \label{3e26}
\widetilde g_0'\ =\ \gamma \widetilde g_0\ ,
\end{equation}
where $\widetilde g_0$ corresponds to solution (\ref{3e20}) and
$\gamma $ is a constant, expressible as a series of the coupling
constant. Equation (\ref{3e19}) then becomes, after replacement of
$\widetilde g_0$ by $\widetilde g_0'$:
\begin{equation} \label{3e27}
\widetilde g_0^{-1} \widetilde \psi \ =\ -\gamma \widetilde V
\widetilde \psi \ ,
\end{equation}
with
\begin{equation} \label{3e28}
\widetilde V\ =\ \widetilde T (1-\gamma \widetilde g_0 \widetilde T)^
{-1}\ .
\end{equation}
The presence of the factor $\gamma $ in Eqs. (\ref{3e27}) and 
(\ref{3e28}) leaves enough fredom for accomplishing the desired
cancellation. For the simplicity of notation, we shall no longer
write explicitly the factor $\gamma$, but when necessary shall
mention its presence.
\par
The integral equation satisfied by the Green's function $\widetilde G$ 
also allows one to construct with standard techniques \cite{f,lcl}
the scalar product of states. 
\par
  
\newpage

\section{Perturbation theory calculation of the potential}
\setcounter{equation}{0}

Our main task in the remaining part of this work is the calculation,
in perturbation theory and within a definite  approximation scheme, 
of potential $\widetilde V$ from formula (\ref{3e18}). The perturbation
series of $\widetilde V$ contains, in addition to the usual Feynman
diagrams, other types of diagram, arising from the presence of the
constraint factor $\widetilde g_0$; we shall call these diagrams
``constraint diagrams''; they are obtained from the usual box-ladder 
type diagrams by the replacement of (at least) one pair of fermion
and antifermion (or boson) propagators by the constraint propagator
$\widetilde g$ [Eq. (\ref{3e6})]. The role of these diagrams will be
to cancel the spurious singularities (referred to the potential)
arising from the calculation of the amplitude and to yield a potential
that has most of the required physical properties.
\par
It was already shown in the case of fermions \cite{ms1} that, when
one of the particles becomes infinitely massive, the contributions
of the constraint diagrams cancel those of the ladder and crossed
ladder diagrams, leaving only the contribution of the one-photon
exchange diagram. Thus the wave equation of the remaining particle
reduces in this limit to the Dirac equation in the presence of the
static potential created by the infinitely massive particle.
\par
In the present work we limit ourselves to the evaluation of the ladder
and crossed ladder diagrams in their leading order, neglecting 
radiative corrections. In this approximation the scattering 
amplitude $\widetilde T$ [Eq. (\ref{3e13c})] can be decomposed as:
\begin{equation} \label{4e1}
\widetilde T\ =\ \sum _{n=1}^{\infty} \widetilde T^{(n)}\ ,
\end{equation}
where $\widetilde T^{(n)}$ is the partial amplitude corresponding to $n$
exchanged photons.
\par
Iteration of Eq. (\ref{3e18}) yields for the potential the expansion:
\begin{equation} \label{4e2}
\widetilde V\ \equiv \ \sum _{n=1}^{\infty}\widetilde V^{(n)}\  
=\ \widetilde T
\sum _{p=0}^{\infty} (\widetilde g_0 \widetilde T)^p\ ,
\end{equation}
where $\widetilde V^{(n)}$ is that part of the potential which comes
from $n$ exchanged photons. The first two terms of $\widetilde V$ are:
\begin{eqnarray}
\label{4e3}
\widetilde V^{(1)} &=& \widetilde T^{(1)}\ ,\\
\label{4e4}
\widetilde V^{(2)} &=& \widetilde T^{(2)} + \widetilde T^{(1)}
\widetilde g_0 \widetilde T^{(1)}\ .
\end{eqnarray}
The corresponding diagrams are represented, for the case of fermions,
in Figs. 1 and 2, respectively. [The ingoing particles have now momenta
$p_1$ and $p_2$ and the outgoing ones $p_1'$ and $p_2'$. The constraint 
diagram is represented by a box with a cross in it.]
\par
Our evaluation of the importance of various terms hinges on the infra-red
counting rules of the bound state problem in QED. Thus, if $\alpha$ is
the dimensionless coupling constant squared, $r$ [Eq. (\ref{2e3})]
counts as $O(\alpha ^{-1})$, the transverse momentum $p^T$ as 
$O(\alpha )$, the transferred momentum squared $t$ as $O(\alpha ^2)$,
etc.. Furthermore, we shall evaluate terms with respect to $x$-space;
therefore, expressions computed in momentum space will undergo 
three-dimensional Fourier transformations and will receive additional
$O(\alpha ^3)$ contributions; thus $\alpha /t$ will be counted as 
$O(\alpha ^2)$, $\alpha ^2/\sqrt {(-t)}$ as $O(\alpha ^4)$, etc..
\par
The one-photon exchange contribution, $\widetilde V^{(1)}$, is of order
$\alpha ^2$ and, in three-dimensional $x$-space, is proportional to
$\alpha /r$. As we shall show in Secs. 5 and 6, the two-photon exchange
contributions, corresponding to $\widetilde V^{(2)}$ [Eq. (\ref{4e4})],
yield as a dominant effect an $O(\alpha ^4)$ term, proportional, in
$x$-space to $\alpha ^2/r^2$. It is therefore expected, generalizing
these results, that the $n$-photon exchange contributions yield as
a dominant effect an $O(\alpha ^{2n})$ term, proportional to
$(\alpha /r)^n$. We shall show, in Secs. 7,8 and 9, within an
approximation scheme, that this is indeed the case. Therefore, the sum
of leading terms of each $\widetilde V^{(n)}$ in Eq. (\ref{4e2}) will
result in a local function of $\alpha /r$.
\par
With the above counting rules, there are still nonleading terms, in
$\widetilde V^{(n)}$ ($n>1$), say, that might compete with leading terms of
$\widetilde V^{(n')}$ with $n'>n$. Such terms have, however, a different
structure than a simple power of $\alpha /r$; they are essentially
momentum dependent and cannot be grouped within a function of 
$\alpha /r$; therefore, they will not be considered in the summation
process, although, in a quantitative perturbation theory calculation
their effects, which begin at $O(\alpha ^6)$, should be taken into
account, like those of the radiative corrections.
\par
To have a consistency check of our summation procedure, we make, in
the following sections, parallel calculations of the scalar potential
(with scalar massless photons) in the bosonic and fermionic cases,
where the Feynman diagrams have rather different structures. One
however expects that the spin independent classical part of the
potential in the fermionic case, when brought in an appropriate
representation, should coincide with the bosonic potential. This 
coincidence does occur and provides additional justification for
our summation rules.
\par

\newpage

\section{Two scalar photon exchange contribution in \protect \\ bosonic 
system}
\setcounter{equation}{0}

We consider in this section a system of two spin-0 particles, with
masses $m_1$ and $m_2$, respectively, interacting with scalar (massless)
photons. The corresponding lagrangian density is:
\begin{equation} \label{5e1}
{\cal L}\ =\ \frac {1}{2} \partial _{\mu}A\partial ^{\mu}A + \partial _
{\mu} \phi _1^{\dagger} \partial ^{\mu}\phi_1 - \phi _1^{\dagger}(m_1^2 +
2m_1gA + g^2A^2)\phi _1 + (1\rightarrow 2)\ .
\end{equation}
We have kept analogy with QED and introduced the interactions with the
photon with the substitutions $m_1\rightarrow m_1+gA$ and $m_2
\rightarrow m_2+gA$. We define:
\begin{equation} \label{5e2}
\alpha \ =\ \frac {g^2}{4\pi}.
\end{equation}
Furthermore, we ignore other types of interaction like $\phi ^4$ or
$A^4$ which arise from renormalization. We assume that the 
corresponding renormalized coupling constants have been put equal to
zero; more generally, these coupling constants are of order $g^4\sim
\alpha ^2$ and the corresponding Feynman diagrams yield contact 
potentials that are of order $\alpha ^5$ and hence can be neglected 
in comparison to $O(\alpha ^4)$ effects.
\par
Our definitions of the propagators are:
\begin{equation} \label{5e3}
G_1(p_1)\ =\ \frac {i}{p_1^2-m_1^2+i\epsilon}\ ,\ \ \ \
G_2(p_2)\ =\ \frac {i}{p_2^2-m_2^2+i\epsilon}\ ,\ \ \ \ 
D(k)\ =\ \frac {i}{k^2+i\epsilon}\ .
\end{equation}
\par
In the process $(1)+(2)\rightarrow (1')+(2')$ we designate the 
corresponding momenta by $p_1$, $p_2$, $p_1'$ and $p_2'$, respectively,
and introduce the usual momentum variables:
\begin{eqnarray} \label{5e4}
s &=& (p_1+p_2)^2\ =\ (p_1'+p_2')^2\ =\ P^2\ =\ P_L^2\ ,\nonumber \\
q &=& p_1-p_1'\ =\ p_2'-p_2\ ,\ \ \ t\ =\ q^2\ ,\ \ \ u\ =\ (p_1-p_2')
^2\ .
\end{eqnarray}
\par
When constraint (\ref{2e8}) is imposed on the external particles, one
obtains:
\begin{equation} \label{5e5}
p_{1L}\ =\ p_{1L}'\ ,\ \ \ p_{2L}\ =\ p_{2L}'\ ,\ \ \ q_L\ =\ 0\ ,
\ \ \ q^2\ =\ q^{T2}\ ,
\end{equation}
with $p_{1L}$ and $p_{2L}$ given by Eqs. (\ref{2e9}).
\par
The Feynman diagrams will be calculated with the external particles
considered off the mass shell, with longitudinal momenta fixed by the
bound state mass $P_L$, through relations (\ref{2e9}), while the
transverse momenta squared $p^{T2}$ and $p^{\prime T2}$ have, according
to our counting rules, orders of magnitude of $\alpha ^2$. The off-mass
shell deviations are represented by the quantities
\begin{equation} \label{5e6}
\lambda ^2\ =\ m_1^2-p_1^2\ =\ m_2^2-p_2^2\ , \ \ \ \ \ \ 
\lambda^{\prime 2}\ =\ m_1^2-p_1^{\prime 2}\ =\ m_2^2-p_2^{\prime 2}\ ,    
\end{equation}
which are positive and are of order $\alpha ^2$. The mean values of 
$p^{T2}$ and $p^{\prime T2}$ in the bound state are equal and hence we
have a similar equality for $\lambda ^2$ and $\lambda^{\prime 2}$:
\begin{equation} \label{5e7}
<p^{T2}>\ =\ <p^{\prime T2}>\ ,\ \ \ \ \ \ <\lambda ^2>\ =\ 
<\lambda^{\prime 2}>\ .
\end{equation}
To the accuracy of the present calculations, where we are evaluating
only leading effects of a set of Feynman diagrams with a fixed number
of photons, we shall not distinguish between $p^{T2}$ and 
$p^{\prime T2}$, and between $\lambda ^2$ and $\lambda^{\prime 2}$, 
and shall use the equalities (\ref{5e7}) without the mean values.
\par
We also introduce the off-mass shell analogue of the quantity $b_0^2(s)$
[Eq. (\ref{3e22})]:
\begin{equation} \label{5e8}
b^2(s)\ =\ \frac {s}{4} - \frac {1}{2} (p_1^2+p_2^2) + \frac {(p_1^2-
p_2^2)^2}{4s}\ =\ -p^{T2}\ ,
\end{equation}
the second equality arising from the use of Eqs. (\ref{2e4}), (\ref{2e8}),
and (\ref{2e9}) and of the definitions $p_1^2=p_{1L}^2+p^{T2}$,
$p_2^2=p_{2L}^2+p^{T2}$. One also easily shows the relations:
\begin{equation} \label{5e9}
b_0^2\ =\ p_{1L}^2-m_1^2\ =\ p_{2L}^2-m_2^2\ =\ -\lambda ^2+b^2\ .
\end{equation}
Notice that $b^2>0$, while $b_0^2<0$. Keeping only the leading order 
terms in $\alpha $, one also has the mean value equality:
\begin{equation} \label{5e10a}
<\lambda ^2>\ =\ 2<b^2>\ ,   
\end{equation}
which in turn implies the leading order equality:
\begin{equation} \label{5e10b}
(p^T+{p'}^T)^2\ =\ -2(\lambda ^2+\frac {t}{2})\ .
\end{equation} 
\par
We first calculate the one-photon exchange contribution (Fig. 1).
Because of the constraint conditions [Eqs. (\ref{5e5})], the momentum
carried by the photon is transverse, $q=(q_L=0,q^T)$ [$q^{T2}=-{\bf q^2}$
in the c.m. frame]; the passage to $x$-space is done with the 
three-dimensional Fourier transformation with respect to $q^T$.
Taking into account the definition of $\widetilde T$ [Eq. (\ref{3e13c})],
one finds for the potential $V^{(1)}$ [Eq. (\ref{4e3})] (the potential
in the bosonic case is not tilded):
\begin{equation} \label{5e11a}
V^{(1)}\ =\ \frac {2m_1m_2}{P_L} \frac {g^2}{t}\ ,
\end{equation}
or, in $x$-space:
\begin{equation} \label{5e11b}
V^{(1)}\ =\ -\frac {2m_1m_2}{P_L} \frac {\alpha }{r}\ .
\end{equation}
\par
We next calculate the two-photon exchange contributions. In addition
to the diagrams of Fig. 2, one also has the two diagrams of Fig. 3.
We first introduce the notations for the various integrals we shall meet
during the calculations. They are similar to those introduced by
Brown and Feynman \cite{bf} and Redhead \cite{r} (but our volume 
element $d^4k$ does not contain a $(2\pi)^{-2}$ factor.) We use the
abbreviated notations:
\begin{eqnarray} \label{5e12}
(0) &=& k^2 + i\epsilon\ ,\ \ \ \ (3)\ =\ (q-k)^2 + i\epsilon\ ,
\nonumber \\  
(1) &=& (p_1-k)^2 - m_1^2 + i\epsilon\ ,\ \ \ \ (2)\ =\ (p_2+k)^2
- m_2^2 + i\epsilon \ ,\nonumber \\
(-1') &=& (-p_1'-k)^2 - m_1^2 + i\epsilon \ ,\ \ \ \ (-2')\ =\
(-p_2'+k)^2 - m_2^2 + i\epsilon\ ;\nonumber \\
& &
\end{eqnarray}
\begin{eqnarray} \label{5e13}
J_{\ ,\mu,\mu \nu} &=& \int d^4k \frac {(1,k_{\mu},k_{\mu}k_{\nu})}
{(0)(3)(1)(2)}\ ,\nonumber \\
F_{\ ,\mu} &=& \int d^4k \frac {(1,k_{\mu})}{(3)(1)(2)}\ ,\ \ \ \ \ 
H_{\ ,\mu}\ =\ \int d^4k \frac {(1,k_{\mu})}{(0)(1)(2)}\ ,\nonumber \\
G_{\ ,\mu}^{(1)} &=& \int d^4k \frac {(1,k_{\mu})}{(0)(3)(1)}\ ,
\ \ \ \ \ G_{\ ,\mu}^{(2)}\ =\ \int d^4k \frac {(1,k_{\mu})}
{(0)(3)(2)}\ .  \nonumber \\
& &
\end{eqnarray}
\par
For the crossed diagrams, where (2) is replaced by $(-2')$, we use the
abbreviated notations:
\begin{equation} \label{5e14}
J_{\ ,\mu,\mu \nu}(1,-2')\ =\ \int d^4k \frac {(1,k_{\mu},k_{\mu}
k_{\nu})}{(0)(3)(1)(-2')}\ ,\ \ \ \ {\rm etc.}\ . 
\end{equation}
\par
For the integrals resulting from constraint diagrams and according to the
definitions (\ref{3e6}), (\ref{3e20}), (\ref{3e13c}) and (\ref{3e18}),
we have to suppress one of the boson propagator denominator, (1) or (2),
and replace it by $2i\pi \delta (C)/(2P_L)$, the operation being
symmetric with respect to the interchange $(1)\leftrightarrow (2)$:
\begin{eqnarray} \label{5e15}
J_{\ ,\mu,\mu \nu}^C &=& \frac {2i\pi}{2P_L} \int d^4k \delta (k_L)
\frac {(1, k_{\mu}^T, k_{\mu}^T k_{\nu}^T)}{(0)(3)(1)}\ ,\nonumber \\
F_{\ ,\mu}^C &=& \frac {2i\pi}{2P_L} \int d^4k \delta (k_L) \frac
{(1,k_{\mu}^T)}{(3)(1)}\ ,\ \ \ \ \ H_{\ ,\mu}^C\ =\ \frac {2i\pi}
{2P_L} \int d^4k \delta (k_L) \frac {(1,k_{\mu}^T)}{(0)(1)}\ ,
\nonumber \\
G_{\ ,\mu}^C &=& \frac {2i\pi}{2P_L} \int d^4k \delta (k_L) \frac
{(1,k_{\mu}^T)}{(0)(3)}\ .
\end{eqnarray}
\par
Details of the calculations of the above integrals are presented in
Appendices A and B (see also Ref. \cite{rta}). For the integrals of
the diagrams of Fig. 2 we find:
\begin{eqnarray} 
\label{5e16}
J &=& \frac {i\pi ^3}{s} \int _0^1 \frac {d\gamma }{\gamma [\gamma \lambda
^2-(1-\gamma )t]}\ \left (\frac {\lambda ^2} {\gamma s} - 
\frac {b^2}{s}\right )^{-\frac {1}{2}} \nonumber \\ 
& & +\ \frac {2i\pi^2}{st}\bigg [ -\frac {P_L^2}{2p_{1L}p_{2L}}
\ln (-\frac {(\lambda ^2)^2}{st}) + \frac {P_L}{p_{1L}}\ln (\frac
{p_{1L}}{P_L}) + \frac {P_L}{p_{2L}} \ln (\frac {p_{2L}}{P_L}) +
\frac {P_L^2}{p_{1L}p_{2L}} \bigg ] \nonumber \\
& & + O(\alpha ^3 \ln \alpha ^{-1})\ , \nonumber \\
& & \\
\label{5e17}
J(1,-2') &=& -\frac {2i\pi ^2}{st} \bigg [ -\frac {P_L^2}{2p_{1L}
p_{2L}} \ln (-\frac {(\lambda ^2)^2}{st}) + \frac {P_L^2}{p_{1L}
p_{2L}} \nonumber \\
& & - \frac {P_L^2}{p_{1L}(p_{1L}-p_{2L})} \ln (\frac {p_{1L}}{P_L})
+ \frac {P_L^2}{p_{2L}(p_{1L}-p_{2L})} \ln (\frac {p_{2L}}{P_L})
\bigg ] + O(\alpha ^3 \ln \alpha ^{-1})\ , \nonumber \\
& &  \\
\label{5e18}
J^C &=& - \frac {i\pi ^3}{s} \int _0^1 \frac {d\gamma }{\gamma [\gamma
\lambda ^2 - (1-\gamma )t]}\ \left (\frac {\lambda ^2}{\gamma s}-
\frac {b^2}{s}\right )^{-\frac {1}{2}}\ .
\end{eqnarray}
\par
By taking the sum of $J$, $J(1,-2')$ and $J^C$, one finds that $J^C$
cancels the dominant singularity in $J$, while the next-to-leading
singularity of $J$ cancels the leading singularity of $J(1,-2')$. One
obtains:
\begin{equation} \label{5e19}
J+J(1,-2')+J^C\ =\ \frac {2i\pi ^2}{t} \frac {1}{(p_{1L}^2-p_{2L}^2)}
\ln (\frac {p_{1L}^2}{p_{2L}^2}) + O(\alpha ^3 \ln \alpha ^{-1})\ .
\end{equation}
\par
Taking into account the coupling constants and the other multiplicative
factors, the potential $V^{(2)}$ resulting from the above contribution
becomes:
\begin{equation} \label{5e20}
V^{(2)J}\ =\ -\frac {m_1^2 m_2^2}{P_Lt} \frac {g^4}{\pi ^2} \frac {1}
{(m_1^2-m_2^2)} \ln (\frac {m_1^2}{m_2^2}) + O(\alpha ^5 \ln \alpha ^{-1})
\ .
\end{equation}
This is a spurious $O(\alpha ^3)$-term, since the potential is known,
from other semirelativistic calculations, not to possess such a term.
It is a two-body effect and cannot be removed by one-particle
renormalizations. The origin of this term lies in the off-mass shell
extrapolation of the Lippmann$-$Schwinger formula (\ref{3e18}) with
the use of Eq. (\ref{3e20}). We actually also calculated the contribution
of the three diagrams of Fig. 2 on the mass shell, giving to the photon
a small mass. We found that the sum of the three terms of Eq. (\ref{5e19})
vanishes (up to $O(\alpha ^3 \ln \alpha ^{-1})$) and hence the term
(\ref{5e20}) is absent from the potential. If we admit that at leading
orders on-mass shell and off-mass shell calculations should coincide
as far as the potential is concerned, we find another justification
for removing the above term by an appropriate renormalization of the 
Lippmann$-$Schwinger formula. This prescription amounts to requiring
that the $1/r$-terms in the potential arise solely from the one-photon
exchange diagram. We described the corresponding procedure through
Eqs. (\ref{3e26}) to (\ref{3e28}). Choosing to this order
\begin{equation} \label{5e21}
\gamma\ =\ 1+\gamma _1\ , \ \ \ \ \ 
\gamma _1\ =\ \frac {g^2}{2\pi^2}\frac {m_1m_2}{(m_1^2-m_2^2)}
\ln (\frac {m_1^2}{m_2^2})\ ,
\end{equation}
one can cancel, with the aid of $V^{(1)}$ [Eq. (\ref{5e11a})], the term
$V^{(2)J}$ [Eq. (\ref{5e20})]:
\begin{equation} \label{5e22}
V^{(2)J} + \gamma _1 V^{(1)}\ =\ 0\ .
\end{equation}
\par
Therefore, the net effect of the three diagrams of Fig. 2 can be 
considered as zero to order $\alpha ^4$ in the potential $V^{(2)}$.
\par
We next evaluate the contributions of the diagrams of Fig. 3. The
corresponding integrals are:
\begin{equation} \label{5e23}
G^{(a)}\ =\ -\frac {i\pi^4}{2p_{aL}\sqrt {-t}} + O(\alpha ^3 \ln
\alpha ^{-1})\ \ \ \ \ \ (a=1,2)\ ,
\end{equation}
and their contribution to the potential $V^{(2)}$ is [we recall that
there is a symmetry factor 2 at the four-line vertex]:
\begin{equation} \label{5e24}
V^{(2)G}\ =\ \frac {1}{8P_L} (\frac {m_1^2}{p_{1L}} + \frac {m_2^2}
{p_{2L}}) \frac {g^4}{\sqrt {-t}} + O(\alpha ^5 \ln \alpha ^{-1})\ .
\end{equation}
\par
In $x$-space, the potential $V^{(2)}$ takes the form:
\begin{equation} \label{5e25}
V^{(2)}\ =\ \frac {1}{P_L} (\frac {m_1^2}{p_{1L}}+\frac {m_2^2}{p_{2L}}) 
\frac {\alpha ^2}{r^2}\ ,
\end{equation}
and, together with $V^{(1)}$ [Eq. (\ref{5e11b})], the potential becomes
to order $\alpha^4$:
\begin{equation} \label{5e26}
V\ =\ V^{(1)} + V^{(2)}\ =\ -\frac {2m_1m_2}{P_L} \frac {\alpha }{r}
+ \frac {1}{P_L} (\frac {m_1^2}{p_{1L}}+\frac {m_2^2}{p_{2L}})
\frac {\alpha^2}{r^2}\ .
\end{equation}
\par
In summary, the constraint diagram has cancelled the dominant spurious
logarithmic singularity of the direct diagram, while the remaining
subdominant logarithmic singularities have been cancelled among the 
direct and crossed diagrams. Among the two types of diagram we met in
Figs. 2 and 3, it is only the diagrams of the type of Fig. 3 (the
``seagull'' diagrams) that have contributed to the $O(\alpha ^4)$ effects.
\par
Similar calculations can also be repeated with scalar QED (vector
photons). We did such calculations in an arbitrary covariant linear 
gauge (characterized by a parameter $\xi$). The vector nature of the
interaction renders the calculations more complicated. Furthermore,
the constraint diagram has an ultra-violet divergence and necessitates
an appropriate subtraction. The final qualitative results remain, 
however, the same as above: the spurious singularities disappear
and the leading term is of order $\alpha^4$, with the correct 
coefficient as expected from scalar QED spectroscopy \cite{t1,n} and
the Breit hamiltonian for bosons.
We shall not present the details of these calculations in this article
and shall be content with the relatively simple case of scalar 
interactions for the bosonic case. 
\par   

\newpage

\section{Scalar and vector interactions with fermions}
\setcounter{equation}{0}

We turn in this section to the calculation of the potentials resulting 
from one- and two-sacalar or vector photon exchanges between a fermion
with mass $m_1$ and an antifermion with mass $m_2$. The interaction 
lagrangian densities are
\begin{equation} \label{6e1} 
{\cal L}_{SI}\ =\ g\overline \psi _1 A\psi _1 + (1\rightarrow 2)\ ,
\ \ \ \ \ {\cal L}_{VI}\ =\ e\psi _1 \gamma _{\mu}A^{\mu}\psi _1 +
(1\rightarrow 2)\ ,
\end{equation}
respectively, with the common notation for the coupling constant
squared, Eq. (\ref{5e2}) and
\begin{equation} \label{6e3}
\alpha \ =\ \frac {e^2}{4\pi}\ .
\end{equation}
The fermion propagators are:
\begin{equation} \label{6e4}
S_1(p_1)\ =\ \frac {i}{\gamma _1.p_1-m_1^2+i\epsilon}\ ,\ \ \ \ \ 
S_2(-p_2)\ =\ \frac {i}{-\gamma _2.p_2-m_2^2+i\epsilon}\ ,
\end{equation}
respectively; the scalar photon propagator is given in Eqs. (\ref{5e3}),
while the vector photon propagator is, in the linear gauge 
characterized by a parameter $\xi$:
\begin{equation} \label{6e4c}
D_{\mu\nu}(k)\ =\ -(g_{\mu\nu}-\xi \frac {k_{\mu}k_{\nu}}{k^2})
\frac {i}{k^2+i\epsilon}\ .
\end{equation}
The constraint factor $\widetilde g_0$ [Eqs. (\ref{3e6}), (\ref{3e2}),
(\ref{3e20}) and (\ref{2e11})] is:
\begin{equation} \label{6e5}
\widetilde g_0\ =\ S_1(p_1) S_2(-p_2) H_0 \big \vert _{C(p)}\ .
\end{equation}
\par
The potential $\widetilde V^{(1)}$ that results from one scalar photon
exchange is:
\begin{equation} \label{6e6}
\widetilde V_S^{(1)}\ =\ \frac {g^2}{2P_L t}
\end{equation}
[$t=q^{T2}$, when constraint $C$ is used on the external particles,
Eqs. (\ref{5e5})], which becomes in $x$-space:
\begin{equation} \label{6e7}
\widetilde V_S^{(1)}\ =\ -\frac {\alpha }{2P_L r}\ .
\end{equation}
\par
For the vector interaction one obtains for $\widetilde V^{(1)}$:
\begin{equation} \label{6e8}
\widetilde V_V^{(1)}\ =\ -\frac {e^2}{2P_L}\gamma _1^{\mu}
\gamma _2^{\nu} (g_{\mu\nu}-\xi \frac {q_{\mu}^T q_{\nu}^T}{t}) \frac {1}
{t}\ ,
\end{equation}
which becomes in $x$-space:
\begin{equation} \label{6e9}
\widetilde V_V^{(1)}\ =\ \frac {\alpha }{2P_L r} \gamma _1^{\mu} \gamma
_2^{\nu} (g_{\mu\nu}-g_{\mu\nu}^{TT} \frac {\xi}{2} + \frac
{x_{\mu}^Tx_{\nu}^T}{x^{T2}} \frac {\xi}{2})\ .
\end{equation}
Upon comparing this expression with Eq. (\ref{2e26}) [notice that the
difference between $\widetilde V$ and $V$ [Eq. (\ref{2e22})] shows
up only at third formal order in $\alpha$] one identifies the vector
potentials $V_2$, $U_4$ and $T_4$ in lowest order:
\begin{equation} \label{6e10}
V_2\ =\ \frac {\alpha }{2P_Lr}\ ,\ \ \ \ \ U_4\ =\ (1-\frac {\xi}{2})
\frac {\alpha }{2P_Lr}\ ,\ \ \ \ \ T_4\ =\ \frac {\xi}{2} \frac 
{\alpha}{2P_Lr}\ .
\end{equation}
In the Feynman gauge ($\xi =0$) one observes the simple relations:
\begin{equation} \label{6e11}
V_2\ =\ U_4\ ,\ \ \ \ \ T_4\ =\ 0\ .
\end{equation}
\par
The difference between the timelike vector potential $V_2$ and the 
spacelike vector potentials $U_4$ and $T_4$ is that, at each given formal
order in $\alpha$, the latter, because of the multiplicative
$\gamma ^T$-matrices, contribute to terms that are $O(\alpha ^2)$ times
less important than the contribution of the former. This feature makes
the computation of the potentials $U_4$ and $T_4$ rather tricky, since 
they contribute only to nonleading terms, while $V_2$ contributes to
the leading ones. In the present work we shall only calculate $V_2$.
However, if one makes the assumption that relations (\ref{6e11}) obtained
in lowest order in the Feynman gauge remain valid in higher orders, the
knowledge of $V_2$ will allow the reconstitution of the whole vector
potential in this gauge.
\par
For the calculation of $\widetilde V^{(2)}$ we need to calculate the
contributions of the three diagrams of Fig. 2. To this end, we 
transform the fermion propagators by bringing the $\gamma$-matrices
in the numerators and giving the denominators a bosonic form. Notice
that the passage from the direct to the crossed diagram is obtained
by the replacement $p_2+k\rightarrow p_2'-k$, which, in the bosonic
denominators is equivalent to $(-p_2'+k)$. We then obtain for the
scalar potential $\widetilde V_S^{(2)}$ the following decomposition
in terms of the integrals (\ref{5e13})-(\ref{5e15}):
\begin{eqnarray} \label{6e13}
\widetilde V_S^{(2)} \left (\frac {i}{2P_L} \frac {g^4}{(2\pi)^4}
\right )^{-1} &=& (\gamma _1.p_1+m_1)(-\gamma _2.p_2+m_2)(J+J(1,-2')
+J^C) \nonumber \\
& & -(\gamma _1.p_1+m_1)\gamma _2^{\mu} (J_{\mu}-J_{\mu}(1,-2')+
J_{\mu}^C) \nonumber \\
& & -(-\gamma _2.p_2+m_2)\gamma _1^{\mu} (J_{\mu}+J_{\mu}(1,-2')+
J_{\mu}^C) \nonumber \\
& & +\gamma _1^{\mu}\gamma _2^{\nu} (J_{\mu \nu}-J_{\mu \nu}(1,-2')
+J_{\mu \nu}^C) \nonumber \\
& & -\big [(\gamma _1.p_1 + m_1) J(1,-2') - \gamma _1.J(1,-2')
\big ]\gamma _2.q\ .
\end{eqnarray}
The transverse matrices $\gamma ^T$ are counted as being of order
$\alpha$, since they connect nondominant components of the wave
function to the dominant one. We summarize here the leading 
contributions in $x$-space of the integrals of Eq. (\ref{6e13}) 
(see Appendices A and B):
\begin{eqnarray} \label{6e14}
& & J+J(1,-2')+J^C\ =\ O(\alpha ^3 \ln \alpha ^{-1})\ , \nonumber \\
& & J_L\ =\ \frac {1}{2P_L} (G^{(1)}-G^{(2)})\ ,\ \ \ \ \
J_L(1,-2')\ =\ -\frac {1}{2P_L} (G^{(1)}+G^{(2)})\ , \nonumber \\
& &\gamma _2^T.(J^T-J^T(1,-2')+J^C)\ =\ O(\alpha ^3 \ln \alpha ^{-1})\ ,
\nonumber \\
& &\gamma _1^T.(J^T+J^T(1,-2')+J^C)\ =\ O(\alpha ^3 \ln \alpha ^{-1})\ ,
\nonumber \\
& &\gamma _1^{\mu} \gamma _2^{\nu} (J_{\mu \nu}-J_{\mu \nu}(1,-2')+
J_{\mu \nu}^C)\ =\ O(\alpha ^3 \ln \alpha ^{-1})\ , \nonumber \\
& & \big [(\gamma _1.p_1+m_1) J(1,-2') - \gamma _1.J(1,-2')\big ]
\gamma _2.q\ =\ O(\alpha ^3 \ln \alpha ^{-1})\ .\nonumber \\ 
& &
\end{eqnarray}
The first relation above is understood in the sense of Eq. (\ref{5e22}).
The expressions of $G^{(1)}$ and $G^{(2)}$ are given in Eqs. (\ref{5e23}).
\par
After replacing the matrices $\gamma _{1L}$ and $\gamma _{2L}$ by
their eigenvalues $+1$ and $-1$, respectively, with respect to the
dominant component of the wave function and making the approximations
$m_1\approx p_{1L}$, $m_2\approx p_{2L}$, one finally obtains:
\begin{equation} \label{6e16}
\widetilde V_S^{(2)}\ =\ \frac {g^4}{16P_L^2} \frac {1}{\sqrt {-t}}\ ,
\end{equation}
which, in $x$-space, is equivalent to:
\begin{equation} \label{6e17}
\widetilde V_S^{(2)}\ =\ \frac {\alpha ^2}{2P_L^2 r^2}\ .
\end{equation}
\par
The scalar potential is then, up to two-photon exchanges:
\begin{equation} \label{6e18}
\widetilde V_S\ =\ \widetilde V_S^{(1)} + \widetilde V_S^{(2)}\ =\
-\frac {\alpha }{2P_Lr} + \frac {\alpha ^2}{2P_L^2 r^2}\ .
\end{equation}
\par
For the vector potential one has to add the $\gamma $-matrices at the
vertices. For the direct and constraint diagrams one obtains 
expressions having the following structure: $\gamma _1.\gamma _2$
$[\ \ ]$ $\gamma _1.\gamma _2$, where the quantity $[\ \ ]$ is the
corresponding expression obtained in the scalar interaction case.
For the crossed diagram the structure is: $\gamma _{1\alpha}\gamma _2
^{\beta}$ $[\ \ ]$ $\gamma _{1\beta} \gamma _2^{\alpha}$, with $[\ \ ]$
representing the corresponding expression of the scalar case. The 
interchange of the matrices $\gamma _2^{\beta}$ and $\gamma _2^{\alpha}$
brings in new terms, which, however, turn out to be of order
$\alpha ^3 \ln \alpha ^{-1}$ compared to the right-hand side of
Eq. (\ref{6e13}). Therefore, to order $\alpha ^4$ in the potential, the 
contribution of the three diagrams can be obtained from the right-hand 
side of Eq. (\ref{6e13}) with the global substitution $\gamma _1.
\gamma _2$ $[\ \ ]$ $\gamma _1.\gamma _2$ and the cancellations
met in the scalar case operate again. At leading order, one finds
for the timelike vector potential:
\begin{equation} \label{6e19}
\widetilde V_V^{(2)}\ =\ -\gamma _{1L}\gamma _{2L} \frac {e^4}
{16P_L^2} \frac {1}{\sqrt {-t}}\ ,
\end{equation}
or, in $x$-space:
\begin{equation} \label{6e20}
\widetilde V_V^{(2)}\ =\ -\gamma _{1L}\gamma _{2L} \frac {\alpha ^2}
{2P_L^2 r^2}\ .
\end{equation}
\par
The timelike vector potential is then, up to two-photon exchanges:
\begin{equation} \label{6e21}
\widetilde V_V\ =\ \widetilde V_V^{(1)}+\widetilde V_V^{(2)}\ =\
\gamma _{1L}\gamma _{2L} (\frac {\alpha }{2P_Lr} - \frac {\alpha ^2}
{2P_L^2 r^2})\ .
\end{equation}
\par
It can be shown \cite{ms1} that when the two-photon exchange 
contribution (\ref{6e20}) is added to the complete one-photon exchange
contribution (\ref{6e10})-(\ref{6e11}) and the nonrelativistic limit
of the wave equations (\ref{2e13a})-(\ref{2e13b}) is taken to order
$1/c^2$, and after a wave function transformation that is equivalent
to the passage from the Feynman gauge to the Coulomb gauge, one 
obtains the Breit hamiltonian \cite{bsblp} which is known to yield
the correct bound state spectrum to order $\alpha ^4$. This is also
an indication that no other $O(\alpha ^4)$-terms should arise from
higher order diagrams. (Verification of correct $O(\alpha ^4)$-terms
with Todorov's electromagnetic potential was presented in Ref. 
\cite{cbwva}.)
\par
Result (\ref{6e17}) was also shown \cite{ms1} to contain the correct
retardation effects, calculated in Refs. \cite{om,bgxz} by differnt
methods.
\par
In summary, the same type of cancellations as for the bosonic case
has occurred for the fermions, despite the fact that the tensor structure 
of the Feynman diagrams is more complicated. The terms that contribute to
the leading [$O(\alpha ^4)$] effects are those coming from the 
longitudinal components of the vector integrals $J_{\mu}$. In these
integrals, the linear contribution of the longitudinal component of
the photon momentum of the numerator cancels the dominant contribution of 
one of the bosonic denominators of the fermion propagators and transforms 
the whole integral into a $G$-type integral (\ref{5e13}). All happens
as if the fermions had effective interactions of the type of Fig. 3.
\par

\newpage

\section{Cancellations with constraint diagrams} 
\setcounter{equation}{0}

The fact that the two-photon exchange diagrams globally produce in
leading order ($\alpha ^4$), both for the bosonic and fermionic cases,
local potentials with the correct coefficients, as compared to the 
bound state spectrum analysis to order $1/c^2$, suggests that such a
result might also hold with $n$-photon exchange ($n>2$) diagrams; in 
this case, we should expect to have a local potential proportional to
$(\alpha /r)^n$. It is, however, not possible to show such a result in
a rigorous way, because of the complexity of the $n$-photon exchange
diagrams. For this reason, it is necessary to device an approximation
scheme that takes into account the properties of the bound state we are
considering, as well as the cancellation mechanisms present in the 
two-photon exchange cases.
\par
The kinematic domain of the bound state is characterized by the fact
that the transverse momenta of the external particles are damped, by
an $O(\alpha )$-factor, with respect to the longitudinal momenta and
the masses. Also, the factor $1/r^2$ of the two-photon exchange potentials
represents in momentum space the three-dimensional convolution (up to
multiplicative coefficients) of the two photon propagators, in which the
longitudinal momentum squared, $k_L^2$, has been neglected; this means that 
the entire $q^2$-dependence of the potential (at leading order) comes from
the photons and not from the fermions. 
Finally, the expressions of the potentials we obtained are independent
of the off-mass shell condition imposed on the bosons and the fermions
(i.e., do not depend on $p_1^2-m_1^2$ or $p_2^2-m_2^2$).
\par
It is then natural to assume that, in quantities that are expected
to be globally free of infra-red singularities, the above properties
also reflect themselves in the internal boson or fermion propagators.
We shall therefore consider the approximation that consists in
linearizing with respect to the internal photon momenta the inverse
boson propagators and in neglecting in them the momentum transfer, as
well as off-mass shell terms. One then obtains for the boson 
propagators the expressions:
\begin{eqnarray} \label{7e2}
G_1(p_1-k) &\simeq & \frac {i}{-2p_1.k+i\epsilon}\ ,\ \ \ \ \ 
G_2(p_2+k)\ \simeq \ \frac {i}{2p_2.k+i\epsilon}\ ,\nonumber \\
G_1(-p_1'-k) &\simeq & \frac {i}{2p_1.k+i\epsilon}\ ,\ \ \ \ \ 
G_2(-p_2'+k)\ \simeq \ \frac {i}{-2p_2.k+i\epsilon}\ .
\end{eqnarray}
\par
This approximation can be considered as a variant of the eikonal
approximation \cite{cw,ls,aibizj} adapted to the bound state problem.
Here, one expects for the components of $k$ the orders of magnitude
\begin{equation} \label{7e3}
|k^T|,\ |k_L|\ =\ O(\alpha)\ . 
\end{equation}
[Actually, $|k_L|$ will come out of the order of $O(\alpha^2)$.]
If the propagators (\ref{7e2}) were multiplied by regular terms, we
could immediately neglect in them $p^T.k^T$, but in many cases the 
multiplicative terms are singular and such an approximation would
be misleading before the isolation of the singularities. Finally,
because of the on-mass shell treatment of the boson or fermion 
propagators in approximations (\ref{7e2}), the photon should be given
a small mass to prevent infra-red divergences from occurring at
intermediary stages.
\par
It can be checked by direct calculation that approximations (\ref{7e2})
yield the correct results (\ref{5e25}), (\ref{6e17}) and (\ref{6e20}).
This can also be checked from the results that we shall obtain for the
general case of $n$-photon exchanges.
\par
We shall show in this section that all types of diagram that have
constraint diagram counterparts are globally cancelled by the latter
in leading order of our approximations. The main formula we use is a
generalization of an identity already used in the eikonal approximation
(cf. Ref. \cite{ls}, Appendix, and Ref. \cite{b}, pp. 116-117).
Let $(c_1,c_2,\cdots ,c_{n+1})$ be a set of $(n+1)$ numbers; it can be
devided into $(n+1)$ independent subsets of $n$ numbers, where in each
subset one of the $c_i$'s $(i=1,\cdots , n+1)$ is missing. We have the
identity:
\begin{equation} \label{8e2}
\sum _{i=1}^{n+1} \sum _{perm} \left [(c_1')^{-1} (c_1'+c_2')^{-1}
\cdots (c_1'+c_2'+\cdots +c_n')^{-1}\right ]_{c_i}\ =\
\left (\sum _{i=1}^{n+1} c_i \right )\big / \left (\prod _{j=1}^{n+1}
c_j \right )\ ,
\end{equation}
where $(c_1',c_2',\cdots ,c_n')$ is a permutation of the subset of $n$
numbers $(c_1,c_2$,$\cdots$,$c_{i-1}$,$c_{i+1}$, $\cdots$,$c_{n+1})$.
\par
Let us consider in the bosonic case an $(n+1)$-photon exchange diagram 
of the type of Fig. 4, which generalizes those of Fig. 2.
We denote by $k_1,\cdots ,k_{n+1}$ the momenta carried
by the photons; for definiteness we number them according to their
departure point on the line of boson 1, from the left to the right. 
To take into account momentum conservation, the factor 
$(2\pi )^4 \delta ^4(\sum _{i=1}^{n+1} k_i - q)$ must be included in the
corresponding integral. Since $q_L=0$, we have $\sum_{i=1}^{n+1} k_{iL}
=0$; but according to the approximations (\ref{7e2}) we may also
neglect $q^T$ in the boson propagators; hence, we have there the
approximation
\begin{equation} \label{8e3}
\sum_{i=1}^{n+1} k_i\ =\ 0\ .
\end{equation}
The total number of the diagrams considered above is $(n+1)!$. In a given
diagram we have $n$ propagators of boson 1 and $n$
propagators of boson 2; only $n$ photon momenta appear on each of the 
boson lines (but not the same in general). The $(n+1)!$ diagrams
can be divided into $(n+1)$ sets, where in each set we have $n!$
permutations of boson 2 propagators containing the same set of $n$ 
photon momenta. With approximations (\ref{7e2}), the sum of all these
propagators has the structure of the left-hand side of Eq. (\ref{8e2}),
with $c_i = (2p_2.k_i+i\epsilon)$; hence, we obtain the factor
\begin{equation} \label{8e4}
I_{2,n+1}\ =\ \frac {{\displaystyle \sum _{i=1}^{n+1} (2p_2.k_i+i\epsilon )}}
{{\displaystyle \prod _{j=1}^{n+1} (2p_2.k_j+i\epsilon )}}\ .  
\end{equation}
Using Eq. (\ref{8e3}), we can rewrite $I_{2,n+1}$ in the form:
\begin{eqnarray} \label{8e5}
I_{2,n+1} &=& \frac {{\displaystyle \sum _{i=1}^n (2p_2.k_i+i\epsilon)}}
{{\displaystyle \prod _{j=1}^n (2p_2.k_j+i\epsilon)}}
\left (\frac {1}{2p_2.k_{n+1}+i\epsilon} + \frac {1}{\sum _{i=1}^n
2p_2.k_i+i\epsilon}\right ) \nonumber \\
&=& I_{2,n}\left (\frac {1}{-\sum _{i=1}^n 2p_2.k_i+i\epsilon} +
\frac {1}{\sum _{i=1}^n 2p_2.k_i+i\epsilon} \right )\ =\
-2i\pi \delta (\sum_{i=1}^n 2p_2.k_i)\ I_{2,n}\ .\nonumber \\
& &
\end{eqnarray}
Repetition of the above operation on $I_{2,n}$, $I_{2,n-1}$,
etc., yields the final result:
\begin{equation} \label{8e6}
I_{2,n+1}
\ =\ (-2i\pi)^{n} \delta (2p_2.k_1) \delta (2p_2.(k_1+k_2))\cdots
\delta (2p_2.\sum _{i=1}^{n} k_i)\ .
\end{equation}
\par
The $\delta $-functions can then be used, upon
integrations on the $k_{iL}$'s, in the boson 1 propagators. 
Denoting by $J^{(n+1)}$ the total contribution of the above diagrams
(without the coupling constants and other multiplicative coefficients),
we obtain:
\begin{eqnarray} \label{8e7}
J^{(n+1)} &=& \left (\frac {-i}{2P_L}\right )^{n}\ 
\int \big [ \prod _{i=1}^{n+1} \frac {d^3k_i^T}{(2\pi )^3}
\frac {1}{(k_i^{T2} - \mu ^2 + i\epsilon)} \big ]\ (2\pi )^3
\delta ^3(\sum _{i=1}^{n+1} k_i^T - q^T) \nonumber \\
& &\ \ \ \ \ \  \times \frac {1}{(-2p^T.k_1^T + i\epsilon )}\
\frac {1}{(-2p^T.(k_1^T+k_2^T) + i\epsilon )}\cdots
\frac {1}{(-2p^T.\sum _{j=1}^{n} k_j^T + i\epsilon )}\ ,
\nonumber \\
& &
\end{eqnarray} 
where $\mu $ is a small mass given to the photon, to avoid infra-red 
divergence. Also, the $\delta $-functions (\ref{8e6}) yield for the
$k_{iL}^2$-terms orders of magnitude of $O(\alpha ^4)$, which accounts
for their omission in front of $O(\alpha ^2)$-terms in    
approximations (\ref{7e2}) and in photon propagators. 
\par
On the other hand, we should also take into account the contributions of 
constraint diagrams that are associated with the diagrams considered
above. From the general formula (\ref{4e2}) we obtain for the potential
corresponding to $n+1$ photon exchanges the expansion:
\begin{equation} \label{8e8}
\widetilde V^{(n+1)}\ =\ \widetilde T^{(n+1)} + \sum _{p=1}^{n}\
\sum _{r_1+\cdots +r_{p+1}=n+1} \widetilde T^{(r_1)} \widetilde g_0
\widetilde T^{(r_2)} \widetilde g_0\cdots \widetilde T^{(r_p)}
\widetilde g_0 \widetilde T^{(r_{p+1})}\ ,\ \ \ n\geq 1\ ,\ r_i\geq 1\ ,
\end{equation}
where, in the generic term of the sum, the constraint factor
$\widetilde g_0$ appears $p$ times. A typical diagram, where $p=2$,
is shown in Fig. 5.
The constraint diagrams corresponding to the diagrams
of the type of Fig. 4 are those for which the
amplitudes $\widetilde T^{(r_i)}$ in Eq. (\ref{8e8}) are also of the
same type, with $r_i$ exchanged photons. Now the longitudinal momentum
transfer between the external particles and a constraint factor or
between two successive constraint factors is zero; furthermore, according
to approximations (\ref{7e2}), we can also neglect transverse momentum
transfers in the bosonic propagators between external particles and
constraint factors and between two successive constraint factors. Hence,
for the partial amplitudes $\widetilde T^{(r_i)}$ ($r_i \geq 2$)
in Eq. (\ref{8e8}) we obtain integrals $J^{(r_i)}$ that have the same
structure as in Eq. (\ref{8e7}), with $n$ replaced by $r_i-1$, except
for the factor $(2\pi )^3 \delta ^3$ which is relative to the whole
diagram. Next, the constraint factor $\widetilde g_0$ provides the
coefficient $(2i\pi )/(2P_L)$ multiplied by the boson propagator
contribution $1/(-2p^T.\sum _{j=1}^{r_1+r_2+\cdots +r_i} k_j^T +
i\epsilon)$, and so forth. Therefore, the constraint diagrams yield
globally the same expressions as $J^{(n+1)}$ [Eq. (\ref{8e7})], except
for a sign factor equal to $(-1)^p$ if $\widetilde g_0$ occurs $p$
times in the expansion (\ref{8e8}). Moreover, the $p$ factors 
$\widetilde g_0$ may appear in 
$\left (_{{\displaystyle  p }}^{{\displaystyle n}}\right )$ 
different configurations. The total contribution of diagrams of the 
type of Fig. 4 and of 
their associated constraint diagrams is then proportional to
\begin{equation} \label{8e9}
J^{(n+1)}\ \big [ 1+\sum _{p=1}^{n} (-1)^p 
\left (_{{\displaystyle p }}^{{\displaystyle n}}
\right ) \big ]\ =\ (1-1)^{n+1}\ =\ 0\ ,\ \ \ \ n\geq 1\ .
\end{equation}
\par 
This result generalizes the one obtained in the two-photon exchange case.
\par
With diagrams involving more than two-photon exchanges, we can 
meet configurations that are mixtures of diagrams of the types of Fig. 4
and of generalizations of Fig. 3. In this case the permutational
procedure on a given line should be applied to those parts of the
diagrams for which the momentum configuration on the other line is
unchanged. Furthermore, to show the cancellation mechanism by
constraint diagrams, not all propagators need be permuted; it is
sufficient to focus attention on those propagators that do not
correspond to the basis of a seagull diagram (of the type of Fig. 3),
since it is only these that may have constraint diagram counterparts.
A simple example illustrates this situation. Let us consider the six
diagrams of Fig. 6. We apply on line 2 the permutational method for the
first two diagrams [(a) and (b)]. The result is a 
$\delta$-function which yields the constraint diagram (a) of Fig. 7 
with an opposite sign and hence a mutual cancellation occurs. Similarly, 
graphs (c) and (d) of Fig. 6 are cancelled by the constraint 
graph (b) of Fig. 7.
\par
The application of the permutations on line 2 in diagrams (e) and
(f) of Fig. 6 yields also a $\delta$-function, but the resulting
expression does not have a constraint diagram counterpart. It is
proportional to:
\begin{equation} \label{8e10}
\delta (p_2.k_2)\ \frac {1}{-p_1.k_1+i\epsilon}\ \frac {1}
{-p_1.(k_1+k_2)+i\epsilon}\ .
\end{equation}
After an integration with respect to $k_{2L}$ this expression becomes
proportional to:
\begin{equation} \label{8e11}
\frac {1}{p^T.k_2^T}\left (\frac {1}{-p_{1L}k_{1L}-p^T.k_1^T+
i\epsilon}-\frac {1}{-p_{1L}k_{1L}-p^T.k_1^T-P_Lp^T.k_2^T/p_{2L}
+i\epsilon}\right )\ .
\end{equation}
Finally, by making in the second term the changes of variable
\begin{equation} \label{8e12}
k_{1L}\rightarrow k_{1L}'=k_{1L}+P_L\frac {p^T.k_2^T}{p_{1L}p_{2L}}\ ,
\ \ \ \ k_{3L}\rightarrow k_{3L}'=k_{3L}-P_L\frac {p^T.k_2^T}
{p_{1L}p_{2L}}\ ,
\end{equation}
the whole integral becomes proportional to:
\begin{equation} \label{8e13}
\frac {1}{p^T.k_2^T} \frac {1}{(-p_{1L}k_{1L}-p^T.k_1^T+i\epsilon)}\
\big [ D(k_{1L}^2+k_1^{T2}) D(k_{3L}^2+k_3^{T2}) - D(k_{1L}^{\prime 2}+
k_1^{T2}) D(k_{3L}^{\prime 2}+k_3^{T2})\big ]\ .
\end{equation}
With the use of assumptions (\ref{7e3}), the difference of the two 
terms inside the brackets yields a non-leading term that can be 
neglected. Therefore, the six diagrams of Fig. 6 are cancelled, at
leading order, by the two constraint diagrams of Fig. 7.
\par
This result can be generalized to more complicated diagrams. Use of the
permutational procedure on a given line produces a certain number of
$\delta$-functions, but only a smaller number of them survives at
leading order; these are precisely those which appear in the constraint
diagrams and are cancelled by them.
\par
We thus end up with the conclusion that the only diagrams that survive
(at leading order) for the calculation of the potential in the bosonic 
case are those which do not have constraint diagram counterparts.
These are generalizations of the diagrams of Fig. 3, some examples of 
which are presented in Figs. 8 and 9.
\par
The above  study can also be applied to the fermionic case. For the 
fermion propagators we use the approximations:
\begin{eqnarray} \label{8e14}
S_1(p_1-k_1) &\simeq  & \frac {i}{-2p_1.k_1+i\epsilon}\ [
(\gamma _{1L}p_{1L}+m_1)-\gamma _{1L}k_{1L} ]\ ,\nonumber \\
S_2(-(p_2+k_2)) &\simeq  & \frac {i}{2p_2.k_2+i\epsilon}\ [
(-\gamma _{2L}p_{2L}+m_2)-\gamma _{2L}k_{2L} ]\ .
\end{eqnarray}
In the numerators we have neglected the transverse momenta $p^T$ and
$k_i^T$ (but not in the denominators); in products of two fermion
propagators we neglect in the numerator the quadratic terms $k_L^2$
(of the same $k_L$). It can be checked that approximations (\ref{8e14})
yield the correct results (\ref{6e17}) and (\ref{6e20}) of the 
two-photon exchange case.
\par
With fermions, we have only diagrams of the type of Fig. 4, but the
presence of the $k_{iL}$'s in the numerators of their propagators gives
rise to a situation that is very similar to that of the bosons. The
products of $(n-1)$ pairs of fermion propagators (for $n$-photon
diagrams) can be decomposed into three types of terms. The first type of
term does not contain any $k_{iL}$ in the numerator. It behaves as in the
bosonic case with diagrams of the type of Fig. 4 and leads to an
expression that is proportional to $J^{(n)}$ [Eq. (\ref{8e7})] and is
cancelled by the corresponding constraint diagram contributions.
In the second type of terms some of the $k_{iL}$'s (but not all)
appear in the numerator. The analysis of these terms is similar
to that of the mixtures of diagrams of the types
of Fig. 3 and Fig. 4 we considered in the bosonic case (but with the 
additional possibility of having more than two photons at
the effective seagull vertices) and their
contributions are again cancelled by corresponding contributions of
constraint diagrams. In the third type of terms products of $(n-1)$
independent combinations of the $k_{iL}$'s appear in the numerator.
These do not have constraint diagram counterparts, since there
$k_{iL}=0$ $(i=1,\cdots ,n)$. Therefore, these terms are the only
surviving parts of the above diagrams for the calculation of the 
potential.
\par

\newpage

\section{The potential in the bosonic case}
\setcounter{equation}{0}

We calculate in this section the scalar potential for two-boson systems.
According to the analysis of Sec. 7 the diagrams that contribute in
higher orders are those of the types of Figs. 8 and 9. The diagrams
of Fig. 8 correspond to an odd number of exchanged photons, while those
of Fig. 9 to an even number of exchanged photons.
\par
We begin by considering diagrams of the type of Fig. 8. Let $(2n+1)$,
with $n\geq 1$, be the number of exchanged photons. The total number
of such diagrams is $(n+1)!(n+1)!$. We denote by $k_1,k_2,$$\cdots$,
$k_{2n+1}$ the momenta carried by the photons, and for definiteness
we number them starting from the photon that leaves the boson line 1,
and then follow continuously the other photons. A global momentum
conservation factor, $(2\pi)^4 \delta^4(\sum _{i=1}^{2n+1} k_i-q)$,
must also be put in the integrals corresponding to these diagrams.
The different diagrams involve permutations of the boson momenta,
which concern, however, independent sets on line 1 and line 2,
respectively. Thus, according to our numbering of photon lines, the
set of internal momenta appearing on line 1 is ($k_1,k_2+k_3,$$\cdots
$,$k_{2n}+k_{2n+1}$), while the set of internal momenta appearing on
line 2 is ($k_1+k_2,$$k_3+k_4,$$\cdots$,$k_{2n-1}+k_{2n},$$k_{2n+1}$).
Furthermore, each set satisfies the equation:
\begin{equation} \label{9e1}
\sum _{i=1}^{2n+1} k_{iL}\ =\ 0\ .
\end{equation}
Since the integrals we are now calculating are of the regular type, we
can, from the start, neglect the transverse momenta $k_i^T$ in the
boson propagators (however, this is not compulsory for the following
calculations). Application of the permutation formulas (\ref{8e2})
and (\ref{8e4})-(\ref{8e6}) for the longitudinal momenta independently
on lines 1 and 2 yields for the boson propagators the result (notice
that formulas (\ref{8e4})-(\ref{8e6}) applied on line 1 with
propagators of the type $(-2p_1.k_i'+i\epsilon)$ still produce 
$(-2i\pi)$-factors):
\begin{equation} \label{9e2}
I_{2n+1}^{(1,2)}\ =\ (-2i\pi)^{2n}\frac {1}{(2p_{1L})^n} \frac {1}
{(2p_{2L})^n} \prod _{i=1}^{2n} \delta (k_{iL})\ ,
\end{equation}
and the corresponding potential becomes:
\begin{eqnarray} \label{9e3}
V^{(2n+1)} &=& \frac {i}{2P_L} (-2im_1g) (-2im_2g) (-2ig^2)^{2n}
i^{2n}i^{2n+1} (-i)^{2n} \frac {1}{(2p_{1L})^n} \frac {1}{(2p_{2L})^n} 
\nonumber \\
& & \times \int \left ( \prod _{i=1}^{2n+1} \frac {d^3k_i^T}{(2\pi)^3}
\frac {1}{(k_i^{T2}+i\epsilon)}\right ) \ (2\pi)^3 \delta ^3
(\sum _{j=1}^{2n+1}k_j^T-q^T)\ ,
\end{eqnarray}
yielding in $x$-space:
\begin{equation} \label{9e4}
V^{(2n+1)}\ =\ -\frac {4m_1m_2}{2P_L} \frac {1}{(p_{1L}p_{2L})^n}
\left (\frac {\alpha }{r}\right )^{2n+1}\ .
\end{equation}
Notice that this formula is also applicable to the case of one-photon
exchange [Eq. (\ref{5e11b})], for which it provides the exact result.
\par
We next consider the diagrams of Fig. 9, with $(2n+2)$ exchanged photons
($n\geq 0$). Here, we have to distinguish between the two types of diagram,
according to whether the photon lines begin and end on line 1 (Fig. 9a)
or on line 2 (Fig.9b). We designate by $V^{(1)(2n+2)}$ and $V^{(2)(2n+2)}$ 
the corresponding potentials, respectively. Let us, for definiteness, 
consider the diagrams of the first type (Fig. 9a). 
The number of such diagrams is
$\frac {1}{2}(n+2)!(n+1)!$, where the factor 1/2 comes from the existing
symmetry with respect to fixed positions of starting and ending photon
lines. The remaining part of the calculation parallels that of $(2n+1)$
exchanged photons. We have on line 1 the set of $(n+2)$ independent
internal momenta ($k_1,$$k_2+k_3,$$\cdots $,$k_{2n}+k_{2n+1},$$k_{2n+2}$)
and on line 2 the set of $(n+1)$ independent momenta
($k_1+k_2,$$k_3+k_4,$$\cdots$,$k_{2n+1}+k_{2n+2}$), each satisfying
Eq. (\ref{9e1}) (with $2n+1$ replaced by $2n+2$). Since the
permutational procedure necessitates $(n+2)!$ permutations on line 1
and $(n+1)!$ permutations on line 2, we have to put the combinatorial
factor 1/2 in front of the final result. We find:
\begin{eqnarray} \label{9e5}
V^{(1)(2n+2)} &=& \frac {i}{2P_L} (-2im_1g)^2 (-2ig^2)^{2n+1} i^{2n+1}
i^{2n+2} (-i)^{2n+1} \frac {1}{2} \frac {1}{(2p_{1L})^{n+1}}
\frac {1}{(2p_{2L})^n}\nonumber \\
& & \times \int \left ( \prod _{i=1}^{2n+2} \frac {d^3k_i^T}{(2\pi)^3}
\frac {1}{(k_i^{T2}+i\epsilon)}\right )\ (2\pi)^3 \delta ^3 
(\sum _{j=1}^{2n+2} k_j^T-q^T)\ ,\nonumber \\
& &
\end{eqnarray}
yielding in $x$-space:
\begin{eqnarray} \label{9e6}
V^{(1)(2n+2)} &=& \frac {4m_1^2p_{2L}}{4P_L} \left (\frac {\alpha }
{\sqrt {p_{1L}p_{2L}} r}\right )^{2n+2}\ .\nonumber \\
& &
\end{eqnarray}
$V^{(2)(2n+2)}$ is obtained from Eq. (\ref{9e6}) with the exchanges
$1\leftrightarrow 2$.
\par
Collecting all the contributions calculated above, we obtain for the
total potential:
\begin{eqnarray} \label{9e7}
V &=& -\frac {2m_1m_2}{P_L}\ \frac {\alpha }{r}\ \sum _{n=0}^{\infty}
\left (\frac {\alpha }{\sqrt {p_{1L}p_{2L}} r} \right )^{2n}
\nonumber \\
& & + \frac {(m_1^2p_{2L}+m_2^2p_{1L})}{P_L} \left (\frac {\alpha }
{\sqrt {p_{1L}p_{2L}} r}\right )^2 \sum _{n=0}^{\infty} \left (
\frac {\alpha }{\sqrt {p_{1L}p_{2L}} r}\right )^{2n}\ ,
\end{eqnarray}
which on summation becomes:
\begin{eqnarray} \label{9e8}
V &=& -\frac {2m_1m_2}{P_L} \frac {\alpha }{r} + 
\left (\frac {m_1^2p_{2L}+m_2^2p_{1L}}{P_L}-\frac {2m_1m_2}{P_L}
\frac {\alpha }{r}\right ) 
\left (\frac {\alpha }{\sqrt {p_{1L}p_{2L}} r}\right )^2 \big /
\left (1- (\frac {\alpha }{\sqrt {p_{1L}p_{2L}} r})^2\right ).
\nonumber \\
& &
\end{eqnarray}
\par
This expression can be further simplified. We used for the evaluation
of higher order diagrams approximations where we neglected (in the
regular expressions) the transverse momenta of the external particles
(as they contribute to nonleading terms). This is equivalent to using the
static limit in these diagrams. Therefore we are entitled to use the
equations of motion in the static limit in the corresponding expressions
to eliminate energy factors in terms of the masses and energy independent
potentials. Using the equations of motion (\ref{2e1a})-(\ref{2e1b}) in
the static limit ($p^{T2}=0$) with $V$ given by Eq. (\ref{9e8}), we can,
after a few algebraic manipulations, reexpress the second piece of $V$
(which involves higher order terms only) in terms of simpler quantities. 
The final result for $V$ is:
\begin{equation} \label{9e14}
V\ =\ -\frac {2m_1m_2}{P_L}\frac {\alpha }{r} + \frac {\alpha ^2}{r^2}\ .
\end{equation}
\par
Potential $V$ above has the same structure as the scalar potential proposed
by Crater and Van Alstine \cite{cva1} on the basis of an extension of the
minimal substitution rules introduced by Todorov \cite{t1}, who 
identified the two-particle motion in the c.m. frame to that of a
fictitious particle with reduced mass and energy defined, respectively,
as:
\begin{equation} \label{9e15}
m_W\ =\ \frac {m_1m_2}{P_L}\ ,\ \ \ \ \
E\ =\ (m_W^2+b_0^2)^{1/2}\ =\ \frac {1}{2P_L} (P^2-m_1^2-m_2^2)\ ,
\end{equation}
where $b_0^2$ [Eq. (\ref{3e22})] is the mass-shell invariant relative
momentum squared, corresponding to the c.m. total energy $\sqrt s=P_L$.
It can be checked, after rewriting the Klein$-$Gordon operator $H_0$
[Eq. (\ref{2e11})] in the form
\begin{equation} \label{9e19}
H_0\ =\ E^2-m_W^2+p^{T2}\ ,
\end{equation}
that potential (\ref{9e14}) is generated in the wave equations 
(\ref{2e1a})-(\ref{2e1b}) by the substitution:
\begin{equation} \label{9e18}
m_W^2\ \rightarrow \ (m_W+S)^2\ ,
\end{equation}
with $S=-\alpha /r$.
\par

\newpage

\section{The potential in the fermionic case}
\setcounter{equation}{0}

We turn in this section to the calculation of the scalar and timelike
vector potentials relative to fermion-antifermion systems. The fermion
propagators are considered in approximations (\ref{8e14}). According to
the analysis, at the end of Sec. 7, of $n$-photon exchange diagrams, the
only surviving terms, at leading order, are those containing in the
numerator products of $(n-1)$ independent combinations of the longitudinal
photon momenta $k_{iL}$. These arise from pairs of fermion and antifermion
propagators containing in their numerator one $k_{iL}$. The permutational
procedure already used in other instances produces $\delta$-functions
which allow each $k_{iL}$ to cancel the denominator of the corresponding 
fermion propagator that did not participate in the permutational operation. 
Hence, there remains at the end the convolution of $n$ photon propagators. 
The combinatorial analysis is much simplified if we use these 
cancellations (with the appropriate coefficients) prior to the 
permutational procedure.
\par
In a diagram with $n$ exchanged photons (Fig. 4) we have several
possibilities to choose the $(n-1)$ independent combinations of $k_{iL}$'s.
Considering for definiteness line 1, we can keep the $(n-1)$ $k_{iL}$
factors in the numerators of the propagators; then we have to remove the
$(n-1)$ $k_{iL}$ factors from the numerators of the propagators of line 2.
We can also keep $(n-2)$ $k_{iL}$ factors on line 1 and keep one $k_{iL}$
on line 2, the choice of the latter being completely fixed by the
complementarity condition to the $(n-2)$ $k_{iL}$'s of line 1: the set
of the $(n-1)$ $k_{iL}$'s globally kept must always form an independent
set of vectors. In general, we have the possibility of keeping $p$
$k_{iL}$'s on line 1 and the complementary $(n-1-p)$ $k_{iL}$'s on
line 2.
\par
Let then $p$ be the number of $k_{iL}$'s kept on line 1. This choice can
be made in $\left (_{{\displaystyle \ \ \ p\ }}^{{\displaystyle n-1}}
\right )$ different ways; however, no
freedom is left for the choice of the set of $(n-1-p)$ complementary
$k_{iL}$'s on line 2. During the permutational procedure, we need
$(p+1)!$ permutations to obtain $p$ $\delta$-functions on line 2 from
the propagators not having $k_{iL}$'s in their numerator; similarly, we
need $(n-p)!$ permutations to produce $(n-p-1)$ $\delta$-functions on
line 1. Taking into account the total number of diagrams with $n$
exchanged photons, which is $n!$, we obtain the combinatorial factor
relative to the above permutational procedure:
\begin{equation} \label{10e1}
C_n\ =\ \sum _{p=0}^{n-1} \left (_{{\displaystyle \ \ \ p\ }}^
{{\displaystyle n-1}}\right ) \frac {n!}
{(p+1)!(n-p)!}\ =\ F(-(n-1),-n;2;1)\ =\ \frac {(2n)!}{(n+1)!n!}\ ,
\end{equation}
where $F(a,b;c;z)$ is the hypergeometric function \cite{be}, the value
of which, for $z=1$, is known in terms of $\Gamma$-functions.
\par
After making the approximations $(\gamma _{1L}p_{1L}+m_1)\simeq 2p_{1L}$
and $(-\gamma _{2L}p_{2L}+m_2)\simeq 2p_{2L}$, the integrations on the
$k_{iL}$'s produce the global factor $((-2i\pi)/(2P_L))^{n-1}$. 
The corresponding potential becomes, in the scalar interaction case:
\begin{eqnarray} \label{10e2}
\widetilde V_S^{(n)} &=& \frac {i}{2P_L} (-ig)^{2n} i^{2(n-1)} i^n
\left (\frac {-i}{2P_L}\right)^{n-1} \frac {(2n)!}{(n+1)!n!}\nonumber \\
& &\times \int \big [\prod _{i=1}^n \frac {d^3k_i^T}{(2\pi)^3} \frac {1}
{(k_i^{T2}+i\epsilon)}\big ]\ (2\pi)^3 \delta ^3 (\sum _{j=1}^n
k_j^T-q)\ ,
\end{eqnarray}
yielding in $x$-space:
\begin{equation} \label{10e3}
\widetilde V_S^{(n)}\ =\ 
\frac {(2n)!}{(n+1)!n!}\left (-\frac {\alpha }{2P_L r}\right )^n\ .
\end{equation}
This expression is also valid for the one-photon exchange case [Eq.
(\ref{6e7})], for which it provides the exact result. The two-photon
exchange result [Eq. (\ref{6e17}) is also reproduced.
\par
The total scalar potential becomes:
\begin{equation} \label{10e5}
\widetilde V_S\ =\ \sum _{n=1}^{\infty} (-1)^n \frac {(2n)!}{(n+1)!n!}
\left (\frac {\alpha }{2P_L r}\right )^n\ =\ \frac {1-\sqrt {1+
\frac {2\alpha }{P_L r}}}{1+\sqrt {1+\frac {2\alpha }{P_L r}}}\ .
\end{equation}
\par
It satisfies the inequality (\ref{2e20}), and hence parametrizations
(\ref{2e22}) and (\ref{2e26}) can be used. One obtains:
\begin{eqnarray} \label{10e6}
\widetilde V_S &=& \tanh V_1\ ,\nonumber \\
V_1 &=& -\frac {1}{4} \ln \left (1+\frac {2\alpha }{P_L r}\right )\ .
\end{eqnarray}
\par
The timelike vector potential is obtained from the scalar one, by
multiplying, for an $n$-photon exchange diagram, $\widetilde V_S^{(n)}$
by $(-1)^n(\gamma _{1L}\gamma _{2L})^n$, the factor $-1$ coming from
the vector photon propagator and the factor $\gamma _{1L}\gamma _{2L}$
from the vertices. Replacing $(n-1)$ factors $\gamma _{1L}\gamma _{2L}$
by their eigenvalue $-1$ at leading order, we end up with the global
factor $-\gamma _{1L}\gamma _{2L}$ with respect to $\widetilde V_S^{(n)}$.
We thus find:
\begin{equation} \label{10e7}
\widetilde V_V^{(n)}\ =\ -\frac {(2n)!}{(n+1)!n!}
\left (-\frac {\alpha }{2P_L r}\right )^n \gamma _{1L}\gamma _{2L}\ ,
\end{equation}
leading for the total potential to the expression:
\begin{equation} \label{10e8}
\widetilde V_V\ =\ -\gamma _{1L}\gamma _{2L}\left ( \frac {1-\sqrt
{1+\frac {2\alpha }{P_L r}}}{1+\sqrt {1+\frac {2\alpha }{P_L r}}}
\right )\ .
\end{equation}
\par
In terms of parametrizations (\ref{2e22}) and (\ref{2e26}) we have:
\begin{eqnarray} \label{10e9}
\widetilde V_V &=& \tanh (\gamma _{1L}\gamma _{2L}V_2)\ ,\nonumber \\
V_2 &=& \frac {1}{4} \ln \left (1+\frac {2\alpha }{P_L r}\right )\ .
\end{eqnarray}
\par
If we adopt for the electromagnetic potential the hypothesis that in the
Feynman gauge relations (\ref{6e11}), obtained in lowest order, remain
also valid in higher orders, we can reconstitute the entire potential:
\begin{eqnarray} \label{10e10}
V_2 &=& U_4\ =\ \frac {1}{4} \ln \left (1+\frac {2\alpha }{P_L r}\right )
\ ,\ \ \ \ \ T_4\ =\ 0\ ,\nonumber \\
\widetilde V_V &=& \tanh (\gamma _1.\gamma _2 V_2)\ .
\end{eqnarray}
\par
This potential is the fermionic generalization of Todorov's potential
\cite{t1}, obtained in two spin-0 particle systems from Eqs. (\ref{9e15})
and (\ref{9e19}) with the substitution rule
\begin{equation} \label{9e17}
E^2\ \rightarrow \ (E+V_0)^2\ ,
\end{equation}
where $V_0=\alpha /r$. Indeed, if we go back to Eq. (\ref{2e29}), which
essentially represents the squared two-body Dirac equation, and replace
there the potentials $V_2$, $U_4$ and $T_4$ by their expressions 
(\ref{10e10}), we find that in the classical part of
this equation the electromagnetic interaction is introduced with the
substitution rule (\ref{9e17}).
\par
As to the scalar potential, it does not correspond in its form
(\ref{10e5})-(\ref{10e6}) to the substitution rule (\ref{9e18}), but can
be made compatible with it if in the higher order terms the classical
static equations of motion are used, as in the bosonic case
[Eq. (\ref{9e14})]. To this end, let us replace $V_1$ by its expression
(\ref{10e6}) in Eq. (\ref{2e29}) and retain in the latter the
classical static (and spin independent) part ($p^T=0$):
\begin{equation} \label{10e11}
\frac {P^2}{4} + \frac {(m_1^2-m_2^2)^2}{4P^2} - \frac {M^2}{4}
e^{{\displaystyle 4V_1}} - \frac {(m_1^2-m_2^2)^2}{4M^2} 
e^{{\displaystyle -4V_1}}\ \approx \ 0\ ,
\end{equation}
from which we deduce the relation
\begin{equation} \label{10e12}
P_L\ \approx \ Me^{{\displaystyle 2V_1}}\ ,
\end{equation}
to be used in higher order terms. Equation (\ref{10e11}) can also be
rewritten in the form:
\begin{eqnarray} \label{10e15}
\frac {P^2}{4} + \frac {(m_1^2-m_2^2)^2}{4P^2} &-& \frac {1}{2}
(m_1^2+m_2^2) - m_1m_2(1-e^{{\displaystyle -4V_1}})\nonumber \\
&-& \frac {M^2}{4} e^{{\displaystyle 4V_1}} (1-e^{{\displaystyle
-4V_1}})^2 \ \approx \ 0\ .
\end{eqnarray}
Using in the last term, which concerns higher order terms than the first,
Eq. (\ref{10e2}) for the factor $M^2e^{{\displaystyle 4V_1}}$ we find:
\begin{equation} \label{10e16}
\frac {P^2}{4} + \frac {(m_1^2-m_2^2)^2}{4P^2} - \frac {1}{2}
(m_1^2+m_2^2) - \frac {2m_1m_2}{P_L}S + S^2\ \approx \ 0\ ,
\end{equation}
where $S$ is defined as:
\begin{equation} \label{10e13}
S\ =\ -\frac {\alpha }{r}\ \ .
\end{equation}
The interaction dependent part in Eq. (\ref{10e16}) has the same form
as in the bosonic case [Eq. (\ref{9e14})], after a similar procedure 
was used there, and corresponds to the substitution rule (\ref{9e18}).
\par
This agreement can be considered as a consistency check for our
approximation scheme, applied to bosons and fermions with different
diagrammatic structures. As one would naturally expect, the classical
parts of the corresponding effective interactions do coincide.
\par
Adopting Eq. (\ref{10e16}) as the final form of the classical part of
the effective interaction, one can reconstruct the new expression of
$V_1$ through the identification of the classical static part of Eq.
(\ref{2e29}) with Eq. (\ref{10e16}). One finds:
\begin{equation} \label{10e17}
V_1\ =\ \frac {1}{2} \ln \left (\frac {1}{M} \big [\ \big (m_1^2+\frac 
{2m_1m_2}{P_L}S+S^2\big )^{\frac{1}{2}} + 
\big (m_2^2+\frac {2m_1m_2}{P_L}S+S^2\big )^{\frac {1}{2}}\
\big ]\right )\ ,
\end{equation}
with $S$ defined in Eq. (\ref{10e13}).
\par
The results obtained thus far can be generalized to include the case of 
a mixture of scalar and vector interactions. The calculations can be
repeated as above, with the difference that every scalar photon
propagator (of the scalar interaction case) is now replaced by an
effective propagator that is the sum of the scalar and vector
propagators (including the couplings at the vertices):
\begin{equation} \label{11e1}
g^2D_S\ \rightarrow \ g^2 D_S + e^2 \gamma _{1L}\gamma _{2L} D_V\ .
\end{equation}
It is this sum which is factorized in the Feynman diagram integrals,
while the fermion propagator part remains unaffected. Concerning the
determination of the tensor nature of the interference terms, the
approximations utilized in the fermionic case (in particular, the
replacements in several places of the matrix products 
$\gamma _{1L}\gamma _{2L}$ by $-1$) do not leave enough predictivity
about this question. It turns out, however, that the analysis of the
same problem in the bosonic case is more predictive and suggestive 
of the representation of the interference terms by effective scalar
interactions. We shall adopt this property also for the fermionic case.
\par
When the summation is done with the effective propagator (\ref{11e1}),
the pure vector potential, which has the form (\ref{10e8})-(\ref{10e9}),
can be isolated from the total potential and $\gamma _{1L}\gamma _{2L}$
replaced in the rest by $-1$. The result is:
\begin{eqnarray} \label{11e2}
\widetilde V &=& \tanh (V_1+\gamma _{1L}\gamma _{2L}V_2)\ ,\nonumber \\
V_2 &=& \frac {1}{4}\ln (1+\frac {2}{P_L}V_0)\ ,\ \ \ \ \ V_0\ =\ 
\frac {\alpha }{r}\ ,\nonumber \\
V_1 &=& -\frac {1}{4}\ln \left (1-\frac {\frac {2}{P_L}S} {1+\frac
{2}{P_L}V_0}\right )\ ,\ \ \ \ \ S\ =\ -\frac {\alpha '}{r}\ ,
\ \ \ \ \alpha '\ =\ \frac {g^2}{4\pi}\ .
\end{eqnarray}
\par
Including then the spacelike potentials $U_4$ and $T_4$ with the 
hypothesis (\ref{10e10}), we obtain:
\begin{eqnarray} \label{11e3}
\widetilde V &=& \tanh (V_1+\gamma _{1L}\gamma _{2L}V_2+\gamma_1^T.
\gamma _2^T U_4)\ , \nonumber \\
U_4 &=& V_2\ =\ \frac {1}{4}\ln \left (1+\frac {2}{P_L}V_0\right )\ .
\end{eqnarray}
\par
We again can transform the scalar potential with the aid of the
classical static equations of motion, as in Eqs.
(\ref{10e11})-(\ref{10e17}). We end up with the following expression
of $V_1$:
\begin{eqnarray} \label{11e7}
V_1 &=& \frac{1}{2}\ln \bigg (\frac {1}{M\sqrt{1+\frac {2}{P_L}V_0}}
\big [\ \big (m_1^2(1+\frac {2}{P_L}V_0)+\frac {2m_1m_2}{P_L}S+S^2\big )^
{\frac {1}{2}}\nonumber \\
& &\ \ \ \ \ \ \ \ +\big (m_2^2(1+\frac{2}{P_L}V_0)+\frac {2m_1m_2}{P_L}S
+S^2\big )^{\frac {1}{2}}\ \big ]\bigg )\ ,\ \ \ V_0=\frac {\alpha }{r}\ ,
\ \ \ S=-\frac {\alpha'}{r}\ ,  
\nonumber \\
& &
\end{eqnarray}
with $V_2$ and $U_4$ given by Eqs. (\ref{11e3}). The classical parts of 
the above potentials satisfy Todorov's minimal substitution rules 
\cite{t1}, defined in Eqs. (\ref{9e15})-(\ref{9e19}), (\ref{9e18}) and 
(\ref{9e17}).
\par

\newpage

\section{Reconstitution of the Bethe$-$Salpeter wave \protect \\
function}
\setcounter{equation}{0}

In this section we shall reconstitute, in the fermionic case, the
Bethe-Salpeter wave function in the framework of the local approximation
that we used throughout this work. The exact reconstitution formula
is given by Eq. (\ref{3e16}), where in the left argument of the first
$T$ the constraint (\ref{2e8}) is not used. A first approximation
consists in replacing the factor $T(1-\widetilde g T)^{-1}$ by potential
$\widetilde V$ [Eq. (\ref{3e18})] in which, however, $p_L$ is no longer
submitted to the constraint (\ref{2e8}). This is easily done by leaving,
in the expression $\widetilde V(t,P_L)$, calculated in the local
approximation, the part $q_L^2$ of $t$($=q_L^2+q^{T2}$) as a free
variable, not fixed at its zero value. More precisely,
$q_L=p_L-p_L'$$=p_L-(m_1^2-m_2^2)/(2P_L)$, where $p_L'$ acts on the
right on $\widetilde \psi$ and hence is submitted to the constraint
(\ref{2e8}). We therefore obtain:
\begin{eqnarray} \label{12e0}
\phi(p) &=& S_1(p_1)S_2(-p_2)\frac {2P_L}{i} \int \frac {d^3p^{\prime T}}
{(2\pi)^3} \widetilde V
\big ((p_L-\frac {(m_1^2-m_2^2)}{2P_L})^2+(p^T-p^{\prime T})^2,P_L\big )
\ \widetilde \psi (p^{\prime T})\ .\nonumber \\
& &
\end{eqnarray}
\par
A further approximation can be imposed by neglecting altogether the
$q_L^2$ dependence of $\widetilde V$ in Eq. (\ref{12e0}). In this case
the wave equation (\ref{3e19}), together with Eq. (\ref{6e5}), can be 
used and one obtains:
\begin{equation} \label{12e1}
\phi \ =\ -S_1 S_2 \frac {2P_L}{i} \big [S_1S_2H_0\big ]_{C(p)}^{-1}
\widetilde \psi \ .
\end{equation}
The relative longitudinal momentum is now contained only in the two
propagators $S_1$ and $S_2$. Formulas (\ref{12e0}) or (\ref{12e1}) can 
be used in the calculations of transition amplitudes, form factors and
decay coupling constants.
\par
Of particular interest is the expression of the Bethe$-$Salpeter wave
function at the zero value of the relative longitudinal
coordinate [it enters in the calculation of decay coupling constants]:
\begin{equation} \label{12e2}
\phi(x_L=0)\ =\ \int \frac {dp_L}{2\pi} \phi (p_L)\ .
\end{equation}
Adopting approximation (\ref{12e1}), integration on $p_L$ concerns
only the propagators $S_1$ and $S_2$. One finds:
\begin{eqnarray} \label{12e3}
\int \frac {dp_L}{2\pi}S_1(p_1)S(-p_2) &=& \frac {i}{4P_LH_0}\big \{
(\gamma _1.p_1+m_1)(-\gamma _2.p_2+m_2)\big (\frac {p_{1L}}{E_1}+\frac
{p_{2L}}{E_2}\big )\nonumber \\
& &-\big (\frac {1}{E_1} - \frac {1}{E_2}\big ) H_0
\big [\gamma _{2L}(\gamma _1.p_1+m_1) + \gamma _{1L}(-\gamma _2.p_2+m_2)
\big ]\nonumber \\
& & + \gamma _{1L}\gamma _{2L}H_0\big (\frac {p_{1L}}{E_1}+\frac {p_{2L}}
{E_2}\big ) \big \}_{C(p)}\ ,
\end{eqnarray}
where $E_1$ and $E_2$ are defined as 
\begin{equation} \label{10e4p}
E_1\ =\ \sqrt{m_1^2-p^{T2}}\ ,\ \ \ \ \ E_2\ =\ \sqrt{m_2^2-p^{T2}}\ ,
\end{equation}
and constraint
(\ref{2e8}) is applied in the right-hand side [$p_{1L}$ and $p_{2L}$
are given by Eqs. (\ref{2e9})].
\par
In the framework of local potential approximation, we can replace the
nonlocal operators $E_1$ and $E_2$ by the eigenvalues $p_{1L}$ and
$p_{2L}$, respectively [Eqs. (\ref{2e9})]. Equations 
(\ref{12e1})-(\ref{12e3}) then yield:
\begin{eqnarray} \label{12e4}
\phi (x_L=0,x^T) &=& \big \{1-\frac {1}{2}\big (\frac {1}{p_{1L}}-
\frac {1}{p_{2L}}\big )\big [\gamma _{2L}(-\gamma _2.p_2-m_2)+
\gamma _{1L}(\gamma _1.p_1-m_1)\big ]\nonumber \\
& & +\gamma _{1L}\gamma _{2L}\frac {1}{H_0} (\gamma _1.p_1-m_1)
(-\gamma _2.p_2-m_2)\big \}\ \widetilde \psi (x^T)\ .
\end{eqnarray}
After the use of the wave equations (\ref{2e13a})-(\ref{2e13b}), Eq.
(\ref{12e4}) becomes:
\begin{eqnarray} \label{12e5}
\phi (x_L=0,x^T) &=& \big \{ 1+\gamma _{1L}\gamma _{2L}\widetilde V
-\frac {1}{2}\big (\frac {1}{p_{1L}}-\frac {1}{p_{2L}}\big ) \big [
\gamma _{2L}(\gamma _1.p_1+m_1)\nonumber \\
& & \ \ \ \ \ \ + \gamma _{1L}(-\gamma _2.p_2+m_2)\big ]\widetilde V 
\big \} \ \widetilde \psi (x^T)\ .
\end{eqnarray}
\par
Finally, using parametrization (\ref{2e22}) and the wave function 
transformation (\ref{2e23}) one obtains:
\begin{eqnarray} \label{12e6}
\phi(x_L=0,x^T) &=& \big \{ e^{{\displaystyle \gamma _{1L}\gamma _{2L}
V}} - \frac {1}{2}\big (\frac {1}{p_{1L}}-\frac {1}{p_{2L}}\big )
\big [\gamma _{2L}(\gamma _1.p_1+m_1)\nonumber \\
& & \ \ \ \ \ \ +\gamma _{1L}(-\gamma _2.p_2+m_2)\big ]\sinh V \big \}
\ \psi (x^T)\ ,
\end{eqnarray}
where $\psi$ is normalized according to Eq. (\ref{2e24}) and $V$ is
of the type (\ref{2e26}) and is a function of $x^T$ and $P_L$.
\par

\newpage

\section{Conclusion}

Constraint theory, applied to two-particle systems,
provides a natural basis for a manifestly covariant three-dimensional
reduction of the Bethe$-$Salpeter equation. The two-body potential
is related, as in the quasipotential approach, to 
the off-mass shell scattering amplitude by means of a Lippmann$-$Schwinger
type equation and is calculable in terms of Feynman diagrams.
Perturbation theory is reorganized with the presence of ``constraint
diagrams'' that appear in the course of the three-dimensional
reduction process.
\par
The two-photon exchange diagrams (for scalar
interactions or vector interactions considered in the Feynman gauge)
are globally free of spurious infrared singularities and yield at
leading order local potentials proportional to $(\alpha /r)^2$, where
$r$ is the three-dimensional relative distance in the c.m. frame; these,
together with the one-photon exchange contribution, produce the correct
bound state spectra to order $\alpha ^4$. The $n$-photon exchange diagram 
contributions were evaluated
in an approximation scheme that is a variant of the eikonal
approximation adapted to the bound state problem; the latter produce
at leading order local potentials proportional to $(\alpha/r)^n$.
The series of leading order potentials were summed, producing 
total potentials that are  functions of $r$ and of the c.m. total energy.
The potentials calculated for vector and scalar interactions are 
equivalent, for their classical parts, to those obtained by 
Todorov \cite{t1} and Crater and Van Alstine \cite{cva1} on the basis
of mimimal substitution rules.
\par 
In summing the series of leading order potentials, ambiguities
might arise from possible incorporation or omission of nonleading
terms. A crucial restriction imposed on such ambiguities comes from an
inequality [(\ref{2e20})] that should satisfy the fermionic potential
in order to ensure the positivity of the norm of the wave function.
The potentials obtained in this work do satisfy the above inequality
and hence do not lead to spurious violations of the positivity of the
norm.
\par
Another question that arises concerns the utility of the potential
formalism itself for the evaluation of the bound state spectrum:
would not the scattering amplitude, calculated in the eikonal
approximation, provide the bound state spectrum through the determination
of the positions of its poles? The accuracy of the evaluation of the
potential in the present work goes beyond that of the conventional
eikonal approximation. For the bosonic case for instance, the latter
sums diagrams of the type of Fig. 4. From the potential theory viewpoint,
the net effect of the contribution of such diagrams is equivalent 
to the one-photon exchange contribution (Fig. 1). It is the summation
of the chains of seagull diagrams of the types of Figs. 3,8 and 9,
not taken into account in conventional eikonal approximation, that
provides the genuine multiphoton-exchange contributions to the potential.
Similarly, in the fermionic case, the contributions isolated in the 
present work for the evaluation of the potential are not considered
in conventional eikonal approximation; futhermore, the latter would 
produce a spin independent bound state spectrum. On the other hand, the
potential evaluated above is reinjected into relativistic wave equations 
that take into account more accurately spin dependent effects. We
conclude that potential formalism is necessary for a detailed probe
of the relativistic bound state spectrum.
\par
The generality of the summation method, in which
the photon propagators, except for the counting rules, do not play any 
active role, leaves open the
possibility of extending the above calculations to other types of
interaction, like those corresponding to the exchanges of massive 
particles or those corresponding to effective confining interactions.
Another domain of application might concern the study of the
gauge dependence problem of the electromagnetic two-body potential.
\par
The fact that the constraint theory wave equations can be reduced
to a single Pauli-Schr\"odinger type equation \cite{ms1}, provides
the complementary basis for their simple applicability to relativistic
bound state problems.
\par

\newpage

\appendix
\section{Calculation of the scalar integrals}
\renewcommand{\theequation}{\Alph{section}.\arabic{equation}}
\setcounter{equation}{0}

In this appendix we calculate the scalar integrals defined in Eqs.
(\ref{5e13})-(\ref{5e15}). They have already been evaluated in the
literature \cite{bf,r}, but generally on the mass shell, with a small
mass given to the photon. Since we are using an off-mass shell
formalism, we present here some details of the calculations. Definitions
of variables and approximations were introduced at the beginning of
Sec. 5, while the counting rules of the orders of magnitude were
explained in Sec. 4; we recall that these are referred to $x$-space.
We use for the calculation of the integrals the Feynman 
parametrization method.
\par
We first consider the integral $J$. After an integration on one of the
parameters, we find:
\begin{equation} \label{ae1}
J\ =\ i\pi^2\int _0^1\int_0^1 \frac {d\gamma d\sigma }
{[\gamma \lambda ^2 - (1-\gamma )t]\big [\lambda ^2 + \gamma 
[\sigma p_1^2 + (1-\sigma )p_2^2 - \sigma (1-\sigma )s]\big ]}\ .
\end{equation}
We next integrate with respect to $\sigma$; the result is an $\arctan$
function, from which we isolate the dominat part by expressing it with
respect to the inverse of its argument. In the nondominant part we
make the change of variable $x = \sqrt{\frac{\lambda^2}{\gamma s}-
\frac {b^2}{s}}$. We find:
\begin{equation} \label{ae3}
J\ =\ j+\widetilde J\ ,
\end{equation}
with
\begin{equation} \label{ae4}
j\ =\ \frac {i\pi ^3}{s} \int _0^1 \frac {d\gamma }{\gamma [\gamma 
(\lambda ^2+t)-t]\sqrt {\frac {\lambda ^2}{\gamma s}-\frac {b^2}{s}}}
\ =\ -\frac {i\pi ^3}{\sqrt {s}}\frac {1}{\sqrt {t(-b_0^2t+(\lambda ^2)^2}}
\ln \left (\frac {1-\sqrt {1-\frac {(\lambda ^2)^2}{b_0^2t}}}
{1+\sqrt {1-\frac {(\lambda ^2)^2}{b_0^2t}}}\right ) 
\end{equation}
and
\begin{eqnarray} \label{ae5}
\widetilde J &=& \frac {2i\pi ^2}{st}\int _{\sqrt{-b_0^2/s}}
^{\infty}\ \frac {dx}{[x^2-((\lambda ^2)^2-b_0^2t)/(st)]}\
\big (\arctan (\frac {x}{\beta _1})+\arctan (\frac {x}{\beta _2})\big )\ ,
\nonumber \\
& &
\end{eqnarray}
where
\begin{equation} \label{ae6}
\beta _1\ =\ \frac {p_{1L}}{P_L}\ ,\ \ \ \ \ \ \beta _2\ =\ \frac {p_{2L}}
{P_L}\ .
\end{equation}
\par
The integration domain of $x$ in Eq. (\ref{ae5}) is separated into two
parts by a point lying in the region of the $x$'s of the order of 
$\alpha^{1/2}$. Appropriate approximations of the integrand in each of the
domains lead to the following expression of $\widetilde J$:
\begin{eqnarray} \label{ae7}
\widetilde J\ =\ \frac {2i\pi^2}{st}\big [&-&\frac {1}{2}\big (\frac {1}
{\beta_1}+\frac {1}{\beta_2}\big )\ln \big (-\frac {(\lambda^2)^2}
{st}\big ) + \big (\frac {1}{\beta_1}\ln\beta_1 + \frac {1}{\beta_2}
\ln\beta_2\big )\nonumber \\
&+&\big (\frac {1}{\beta_1}+\frac {1}{\beta_2}\big )\ \big ] + 
O(\alpha^3\ln\alpha^{-1})\ .
\end{eqnarray}
\par
$J(1,-2')$ is obtained from Eq. (\ref{ae1}) by the replacement of $s$
by $u$ [we are working in the approximations $p_2^{\prime 2}=p_2^2$ and
$\lambda^{\prime 2}=\lambda^2$]. One first integrates with respect to 
$\gamma $:
\begin{eqnarray} \label{ae8}
J(1,-2')\ =\ -\frac {i\pi^2}{t} \int _0^1& &\frac {d\sigma}
{\big [ \sigma p_1^2+(1-\sigma)p_2^2-\sigma (1-\sigma)u + \lambda^2
(\lambda^2+t)/t\big ]}\nonumber \\
& &\times \ln \left (-\frac {[\sigma p_1^2+(1-\sigma)p_2^2-\sigma
(1-\sigma)u+\lambda^2]t}{(\lambda^2)^2}\right )\ .
\end{eqnarray}
In the bound state domain we are considering, we have the following
approximate expressions for $s$ and $u$:
\begin{eqnarray} \label{ae9}
s &=& \big (\sqrt {p_1^2+b^2}+\sqrt {p_2^2+b^2}\big )^2\ \simeq \
\big (\sqrt {p_1^2} + \sqrt {p_2^2}\big )^2 \big (1+\frac {b^2}
{\sqrt {p_1^2 p_2^2}}\big )\ ,\nonumber \\
u &\simeq & \big (\sqrt {p_1^2}-\sqrt {p_2^2}\big )^2 - t - b^2
\frac {(\sqrt {p_1^2}+\sqrt {p_2^2})^2}{\sqrt {p_1^2 p_2^2}}\ ,
\end{eqnarray}
which allow us to write:
\begin{eqnarray} \label{ae10}
\sigma p_1^2+(1-\sigma)p_2^2-\sigma(1-\sigma)u &\simeq& \big (\sqrt 
{p_2^2}+\sigma(\sqrt {p_1^2}-\sqrt{p_2^2})\big)^2\nonumber \\
& &+\sigma(1-\sigma)\big[t+b^2\frac {(\sqrt{p_1^2}+\sqrt{p_2^2})^2}
{\sqrt{p_1^2p_2^2}}\big]\ ;
\end{eqnarray}
the second term in the right-hand side is negligible in front of the
first and this leads to the following expression of $J(1,-2')$:
\begin{eqnarray} \label{ae11}
J(1,-2')&=&-\frac{2i\pi^2}{st}\big[-\frac{1}{2}\big(\frac{1}{\beta_1}
+\frac{1}{\beta_2}\big)\ln\big(-\frac{(\lambda^2)^2}{st}\big)
\nonumber \\
& &-\frac{1}{\beta_1(\beta_1-\beta_2)}\ln\beta_1+\frac{1}{\beta_2
(\beta_1-\beta_2)}\ln\beta_2+\big(\frac{1}{\beta_1}+\frac{1}{\beta_2}
\big)\big] + O(\alpha^3\ln\alpha^{-1})\ .\nonumber \\
& &
\end{eqnarray}
\par
The constraint integral $J^C$, which is three-dimensional, is also
calculated with the Feynman parametrization method. After one 
integration one finds:
\begin{equation} \label{ae12}
J^C\ =\ -\frac {i\pi^3}{s} \int_0^1 \frac {d\gamma}{\gamma [\gamma
\lambda^2-(1-\gamma)t]\sqrt{\frac{\lambda^2}{\gamma s}-\frac{b^2}
{s}}}\ =\ -j\ .
\end{equation}
\par
The structure of the $F$-integrals is very similar to that of the
$J$-integrals (\ref{ae1}) and the same types of approximations are
applied. We obtain:
\begin{eqnarray} 
\label{ae12p}
F &=& f+\widetilde F\ ,\\
\label {ae13}
f &=& -\frac{i\pi^3}{s} \int_0^1 \frac {d\gamma}{\gamma \sqrt {\frac
{\lambda^2}{\gamma s}-\frac{b^2}{s}}}\ =\ -\frac{2i\pi^3}{s}
\sqrt{\frac{s}{b^2}} \arctan\sqrt{\frac{b^2}{b_0^2}}\ ,\\
\label{ae14}
\widetilde F &=& -\frac{2i\pi^2}{s} \int _{\sqrt{-b_0^2/s}}
^{\infty}\ \frac{dx}{(x^2+b^2/s)} [\arctan(\frac{x}{\beta_1})
+\arctan(\frac{x}{\beta_2})]\ ,\nonumber \\
&=&\frac{2i\pi^2}{s}\big[ -\frac{1}{2}\big(\frac{1}{\beta_1}+\frac{1}
{\beta_2}\big)\ln(\frac{\lambda^2}{s}) + \big(\frac{1}{\beta_1}\ln
\beta_1+\frac{1}{\beta_2}\ln\beta_2\big) + \big(\frac{1}{\beta_1}+
\frac{1}{\beta_2}\big) \big] + O(\alpha^5\ln\alpha^{-1})\ ,\nonumber \\
& & 
\end{eqnarray}
\begin{eqnarray}
\label{ae15}
F(1,-2') &=& -i\pi^2 \int_0^1 \frac {d\sigma}{[\sigma p_1^2 +
(1-\sigma)p_2^2-\sigma(1-\sigma)u]}\ \ln \left (\frac 
{\sigma p_1^2+(1-\sigma) p_2^2 - \sigma(1-\sigma)u}{\lambda^2} \right)
\nonumber \\
&=& -\frac{2i\pi^2}{s}\big [ -\frac{1}{2}\big(\frac{1}{\beta_1}+\frac{1}
{\beta_2}\big)\ln (\frac{\lambda^2}{s})-\frac{1}{(\beta_1-\beta_2)}
\big(\frac{1}{\beta_1}\ln\beta_1-\frac{1}{\beta_2}\ln\beta_2\big)\big ]
\nonumber \\
& & \ \ \ \ \ \ \ \ \ \ \ \ \ \ \ \ \ \  + O(\alpha^5\ln\alpha^{-1})\ ,\\
\label{ae16}
F^C &=& \frac{i\pi^3}{s}\int_0^1 \frac {d\gamma}{\gamma \sqrt{\frac
{\lambda^2}{\gamma s}-\frac{b^2}{s}}}\ =\ -f\ .
\end{eqnarray}
\par
Notice that with the approximations $p_1^2=p_1^{\prime 2}$, $p_2^2=
p_2^{\prime 2}$, $\lambda^2=\lambda ^{\prime 2}$ we have used, the 
$F$-integrals are independent of $t$ and hence
\begin{equation} \label{ae17}
H\ =\ F\ ,\ \ \ \ H(1,-2')\ =\ F(1,-2')\ ,\ \ \ \ H^C\ =\ F^C\ .
\end{equation}
\par
The $G$-integrals can be expressed in terms of Euler's dilogarithm or
the Spence function. One finds (without the approximations $p_1^2=
p_1^{\prime 2}$ and $p_2^2=p_2^{\prime 2}$):
\begin{equation} \label{ae20}
G^{(1)}\ =\ -\frac{i\pi^4}{\sqrt {-2t(p_1^2+p_1^{\prime 2})+t^2+(p_1^2-
p_1^{\prime 2})^2}} + O(\alpha^3\ln\alpha^{-1})\ ,
\end{equation}
the dominant part of it being:
\begin{equation} \label{ae21}
G^{(1)}\ \simeq\ -\frac{i\pi^4}{2p_{1L}\sqrt{-t}}\ .
\end{equation}
$G^{(2)}$ is obtained from the above equations by the replacements
$1\leftrightarrow 2$.
\par
$G^{(1)}$ does not have a crossed-diagram counterpart, since it depends
only on the particle 1 momenta. On the other hand, the constraint integral
$G^C$ [Eq. (\ref{5e15})] does not correspond to an explicit
diagram in the expansion (\ref{4e4}), but rather may appear through 
integrals involving fermions or vector interactions of bosons. One 
finds:
\begin{equation} \label{ae22}
G^C\ =\ \frac{2i\pi^3}{2P_L}\int_0^1 \frac {d\gamma}{\sqrt{-\gamma
(1-\gamma)t}}\ =\ \frac{i\pi^4}{P_L\sqrt{-t}}\ .
\end{equation}
\par
We emphasize that the instantaneous approximation in the photon
propagators (setting there $k_L=0$) would not produce the results 
found above. For instance, for the $G$-integral one would find twice
the correct result (\ref{ae21}). A detailed analysis of the integral
shows that the photon propagator poles in $k_L$ contribute with a factor 
$-1/2$ with respect to the contribution of the boson propagator poles.
However, the linearization approximation of the inverse bosonic
propagators, as used starting from Sec. 7, makes the instantaneous 
approximation compatible with the leading order results for the 
potential.
\par

\newpage

\section{Calculation of the vector and tensor integrals}
\setcounter{equation}{0}

Most of the vector and tensor integrals can be calculated by
algebraic decompositions with the aid of the vectors $p_1$, $p_2$,
$q$, etc., in terms of the scalar integrals calculated in Appendix A,
except for those decompositions that might involve infra-red divergences,
in which case a direct calculation is necessary. We present here the
final results [notations are those of Appendix A]:
\begin{eqnarray} 
\label{be1}
F_{\mu} &=& q_{\mu}f-\frac{i\pi^2}{s}(\beta_2p_{1\mu}-\beta_1p_{2\mu}-
q_{\mu})\big [ -\big(\frac{1}{\beta_1}+\frac{1}{\beta_2}\big)\nonumber \\
& & \ \ \ +\pi\sqrt{\frac{s}{b^2}}\big( \frac{\lambda^2}{b^2}\arctan
\sqrt{-\frac{b^2}{b_0^2}}-\sqrt{-\frac{b_0^2}{b^2}}\big )\big ] +
O(\alpha^4\ln\alpha^{-1})\nonumber \\
&\equiv& q_{\mu}f+p_{\mu}^{\prime T}\widetilde H\ ,\\
\label{be2}
H_{\mu} &=& F_{\mu}(q=0)\ =\ p_{\mu}^T\widetilde H \ ,
\end{eqnarray}
\begin{eqnarray}
\label{be3}
F_{\mu}(1,-2') &=& -\frac{i\pi^2}{s}\big \{ p_{1\mu}'\frac{1}{\beta_1
\beta_2}+(p_{2\mu}'-p_{1\mu})\big [ -\frac{1}{(\beta_1-\beta_2)^2}
\ln (\frac{\beta_1}{\beta_2}) + \frac{1}{2\beta_1\beta_2}\nonumber \\ 
& & \ \ \ \ \ \ \ \ \ \ \ \ + \frac{1}{2\beta_1\beta_2}\big(\frac
{\beta_1+\beta_2}{\beta_1-\beta_2}\big)\big ]\big \} 
+O(\alpha^4\ln\alpha^{-1})\ ,\nonumber \\
& & \\
\label{be4}
H_{\mu}(1,-2') &=& \frac{i\pi^2}{s}\big \{-p_{1\mu}\frac{1}{\beta_1
\beta_2} + (p_{1\mu}'-p_{2\mu})\big [ -\frac{1}{(\beta_1-\beta_2)^2}
\ln(\frac{\beta_1}{\beta_2}) + \frac{1}{2\beta_1\beta_2}\nonumber \\ 
& & \ \ \ \ \ \ \ \ \ \ \ \ + \frac{1}{2\beta_1\beta_2}\big(\frac
{\beta_1+\beta_2}{\beta_1-\beta_2}\big)\big ]\big \}
+O(\alpha^4\ln\alpha^{-1})\ ,\nonumber \\
& & 
\end{eqnarray}
\begin{eqnarray}
\label{be5}
F_{\mu}^C &=& q_{\mu}F^C + \frac{i\pi^3}{s}p_{\mu}^{\prime T}\sqrt
{\frac{s}{b^2}}\big(\frac{\lambda^2}{b^2}\arctan \sqrt{-\frac{b^2}
{b_0^2}}-\sqrt{-\frac{b_0^2}{b^2}}\big)+O(\alpha^4\ln\alpha^{-1})
\nonumber \\
&\equiv& q_{\mu}F^C+p_{\mu}^{\prime T}\widetilde H^C\ ,\\
\label{be6}
H_{\mu}^C &=& \frac{i\pi^3}{s} p_{\mu}^T \sqrt{\frac{s}{b^2}} \big(
\frac{\lambda^2}{b^2}\arctan \sqrt{-\frac{b^2}{b_0^2}}-\sqrt{-\frac
{b_0^2}{b^2}}\big) + O(\alpha^4\ln\alpha^{-1})\nonumber \\
&=& p_{\mu}^T \widetilde H^C\ ,\\
\label{be7}
G_{\mu}^{(1)} &=& - \frac{\lambda^2}{(p_1^2+p_1^{\prime 2})}p_{1\mu}
G^{(1)} + \big \{\frac {(\lambda^2+2p_1^2)}{2(p_1^2+p_1^{\prime 2})}
- \frac{\lambda^2(p_1^2-p_1^{\prime 2})}{2t(p_1^2+p_1^{\prime 2})}
\nonumber \\
& & \ \ \ \ \ \ \ \ \ \ \ \ \ \ \ \ + \frac{p_1^2(p_1^2-p_1^{\prime 2})
^2}{2t(p_1^2+p_1^{\prime 2})^2} \big \} q_{\mu} G^{(1)} + O(\alpha^4)
\nonumber \\
&\simeq& \frac{1}{2} q_{\mu} G^{(1)}\ ,\\
\label{be9}
G_{\mu}^C &=& \frac{1}{2} q_{\mu} G^C\ .
\end{eqnarray}
$G_{\mu}^{(2)}$ is obtained from $G_{\mu}^{(1)}$ by the replacements
$p_1\rightarrow -p_2$, $p_1'\rightarrow -p_2'$ and $q\rightarrow q$:
\begin{equation} \label{be8}
G_{\mu}^{(2)} \ \simeq\ \frac{1}{2} q_{\mu} G^{(2)}\ .
\end{equation}
\par
For the $J_{\mu}$-integrals one finds:
\begin{eqnarray} \label{be10}
J_{\mu} &=& P_{\mu}\frac{1}{2P^2} (G^{(1)}-G^{(2)}) + q_{\mu}\frac{1}
{2} [J+\frac{1}{t}(F-H)]\nonumber \\
& & +(p^T+p^{\prime T})_{\mu} \frac {1}{(p^T+p^{\prime T})^2} \big [
-\frac{1}{P_L}(p_{1L}G^{(1)}+p_{2L}G^{(2)})-(\lambda^2+\frac{t}{2})J
+\frac{1}{2}(F+H)\big ] \ ,\nonumber \\
& &
\end{eqnarray}
\begin{eqnarray} \label{be11}
J_{\mu}(1,-2') &=& -P_{\mu}\frac{1}{2P^2}\big [ (G^{(1)}+G^{(2)})+
2(\lambda^2+\frac{t}{2})J(1,-2')-\big( F(1,-2')+H(1,-2')\big)\big ]
\nonumber \\
& &+(p^T+p^{\prime T})_{\mu}\frac{1}{(p^T+p^{\prime T})^2}\big [
\frac{1}{P_L}(p_{1L}G^{(1)}-p_{2L}G^{(2)})\nonumber \\
& &+\frac{(p_{1L}-p_{2L})}{P_L}(\lambda^2+\frac{t}{2})J(1,-2')-
\frac{(p_{1L}-p_{2L})}{P_L}F(1,-2')\big ]\nonumber \\
& &+q_{\mu}\frac{1}{2}\big [ J(1,-2')+\frac{1}{t}\big(F(1,-2')-H(1,-2')
\big) \big ]\ ,
\end{eqnarray}
\begin{eqnarray} \label{be12}
J_{\mu}^C &=& -(p^T+p^{\prime T})_{\mu}\frac{1}{(p^T+p^{\prime T})^2}
\big [ G^C+(\lambda^2+\frac {t}{2})J^C-\frac{1}{2}(F^C+H^C)\big ]
\nonumber \\
& & + q_{\mu}\frac{1}{2} [J^C+\frac{1}{2}(F^C-H^C)]\ .
\end{eqnarray}
\par
In showing Eqs. (\ref{6e14}) for the transverse components of the
$J_{\mu}$-integrals, one uses in some places the leading order equality
(\ref{5e10b}).
\par
For the $J_{\mu \nu}$-integrals it is preferable to first make a
decomposition along the vector integrals already calculated and
$g_{\mu\nu}^{TT}$. For the simplicity of presentation, we use certain
relations previously obtained (such as $F=H$, etc.) and the leading
order equality (\ref{5e10b}), and neglect some nonleading terms. We
find:
\begin{eqnarray} \label{be13}
J_{\mu\nu} &=& (P_{\mu}q_{\nu}+P_{\nu}q_{\mu})\frac{1}{4P^2}(G^{(1)}
-G^{(2)}) + g_{\mu\nu}^{TT}\frac{1}{2}\big[\frac{1}{P_L}(p_{1L}G^{(1)}
+p_{2L}G^{(2)}) + \lambda^2J\big]\nonumber \\
& & - \big((p^T+p^{\prime T})_{\mu}q_{\nu}+(p^T+p^{\prime T})_{\nu}q_{\mu}
\big )\big \{ \frac{1}{(p^T+p^{\prime T})^2} \big [ \frac{1}{2P_L}
(p_{1L}G^{(1)}+p_{2L}G^{(2)})+\frac{1}{2}F\big ] \nonumber \\
& & + \frac{1}{4}J\big \} 
+ \big((p^T+p^{\prime T})_{\mu}P_{\nu}+(p^T+p^{\prime T})_{\nu}
P_{\mu}\big)\frac{1}{4P^2}(G^{(1)}-G^{(2)}) \nonumber \\
& & + (p^T+p^{\prime T})_{\mu} (p^T+p^{\prime T})_{\nu}\frac {1}
{2(p^T+p^{\prime T})^2} \big [ -\frac {2}{P_L}(p_{1L}G^{(1)}+p_{2L}
G^{(2)}) - (2\lambda^2+\frac{t}{2})J\nonumber \\ 
& & + F + \widetilde H\big ] 
+ q_{\mu}q_{\nu}\frac{1}{2t}\big [ (-\lambda^2+\frac{t}{2})J +
f - \widetilde H - \frac{1}{P_L} (p_{1L}G^{(1)}+p_{2L}G^{(2)})\big ]\ ,
\nonumber \\
& &
\end{eqnarray}
[$f$ and $\widetilde H$ are defined in Eq. (\ref{be1}),]
\begin{eqnarray} \label{be14}
J_{\mu\nu}(1,-2') &=& -P_{\mu}\frac{1}{P^2}\big [ \frac{1}{2}
(G_{\nu}^{(1)}+G_{\nu}^{(2)}(-2'))+(\frac{\lambda^2}{2}+t)J_{\nu}(1,-2')
\big ]\nonumber \\
& & - \frac{(p_{1L}-p_{2L})}{2P_L}(p^T+p^{\prime T})_{\mu}J_{\nu}(1,-2')
+ \frac{1}{2}q_{\mu}J_{\nu}(1,-2')\nonumber \\ 
& & + g_{\mu\nu}^{TT}\big (1-\frac
{(p_{1L}-p_{2L})^2}{2P_L^2}\big )F(1,-2') + O(\alpha^3)\ ,\nonumber \\
& &
\end{eqnarray}
\begin{eqnarray} \label{be15}
J_{\mu\nu}^C &=& (p^T+p^{\prime T})_{\mu}(p^T+p^{\prime T})_{\nu}
\frac{1}{2(p^T+p^{\prime T})^2} [ \frac{1}{2}(p^T+p^{\prime T})^2 J^C
- \lambda^2J^C - 2G^C\nonumber \\ 
& & + F^C + \widetilde H^C ]
+\big ( (p^T+p^{\prime T})_{\mu}q_{\nu}+(p^T+p^{\prime T})_{\nu}q_{\mu}
\big)\frac{1}{2(p^T+p^{\prime T})^2} [ \frac{1}{2}(p^T+p^{\prime T})^2
J^C\nonumber \\ 
& &- G^C + F^C]
+q_{\mu}q_{\nu}\frac{1}{2t} [(-\lambda^2+\frac{t}{2})J^C - G^C + F^C -
\widetilde H^C] + g_{\mu\nu}^{TT}\frac{1}{2}(G^C+\lambda^2J^C)\ .
\nonumber \\
& &
\end{eqnarray}
[$\widetilde H^C$ is defined in Eq. (\ref{be5}).]
\par

\newpage

\noindent
{\large {\bf Figures}}
\vspace{1 cm}

\noindent
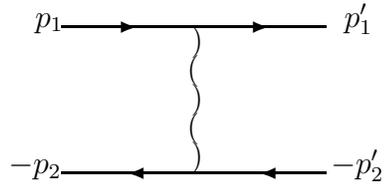
Fig. 1. One-photon exchange diagram.
\vspace{0.3 cm}

\noindent
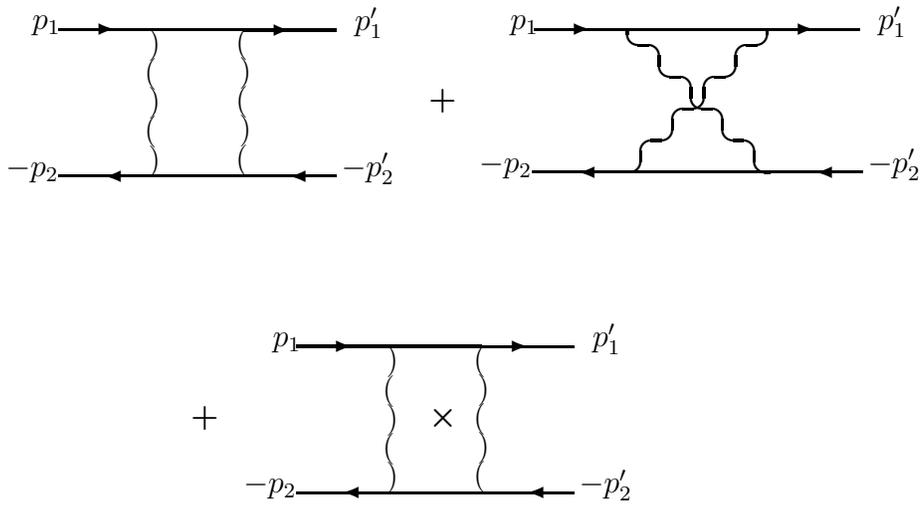
Fig. 2. Two-photon exchange diagrams; the ``constraint diagram'' is
denoted by a cross.
\vspace{0.3 cm}

\noindent
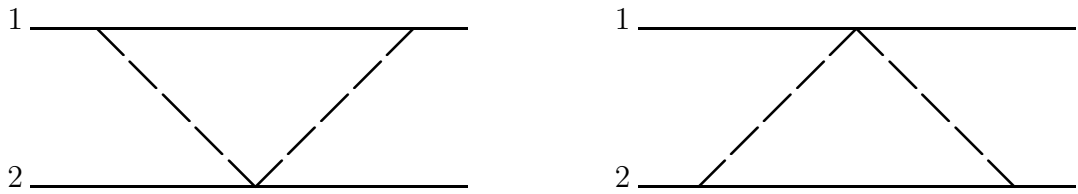
Fig. 3. Two-photon exchange diagrams contributing to the bosonic case.
\vspace{0.3 cm}

\noindent
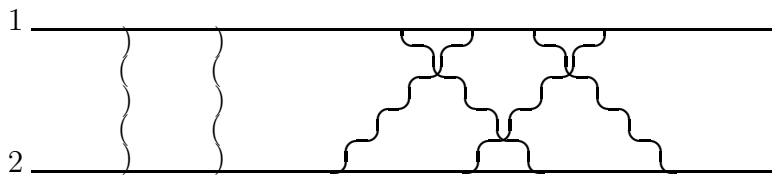
Fig. 4. A multiphoton exchange diagram.
\vspace{0.3 cm}

\noindent
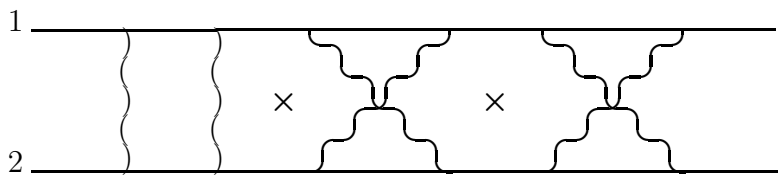
Fig. 5. A typical ``constraint diagram'' where the constraint factor
$\widetilde g_0$ [Eq. (\ref{4e2})] appears twice ($p=2$).
\vspace{0.3 cm}

\noindent
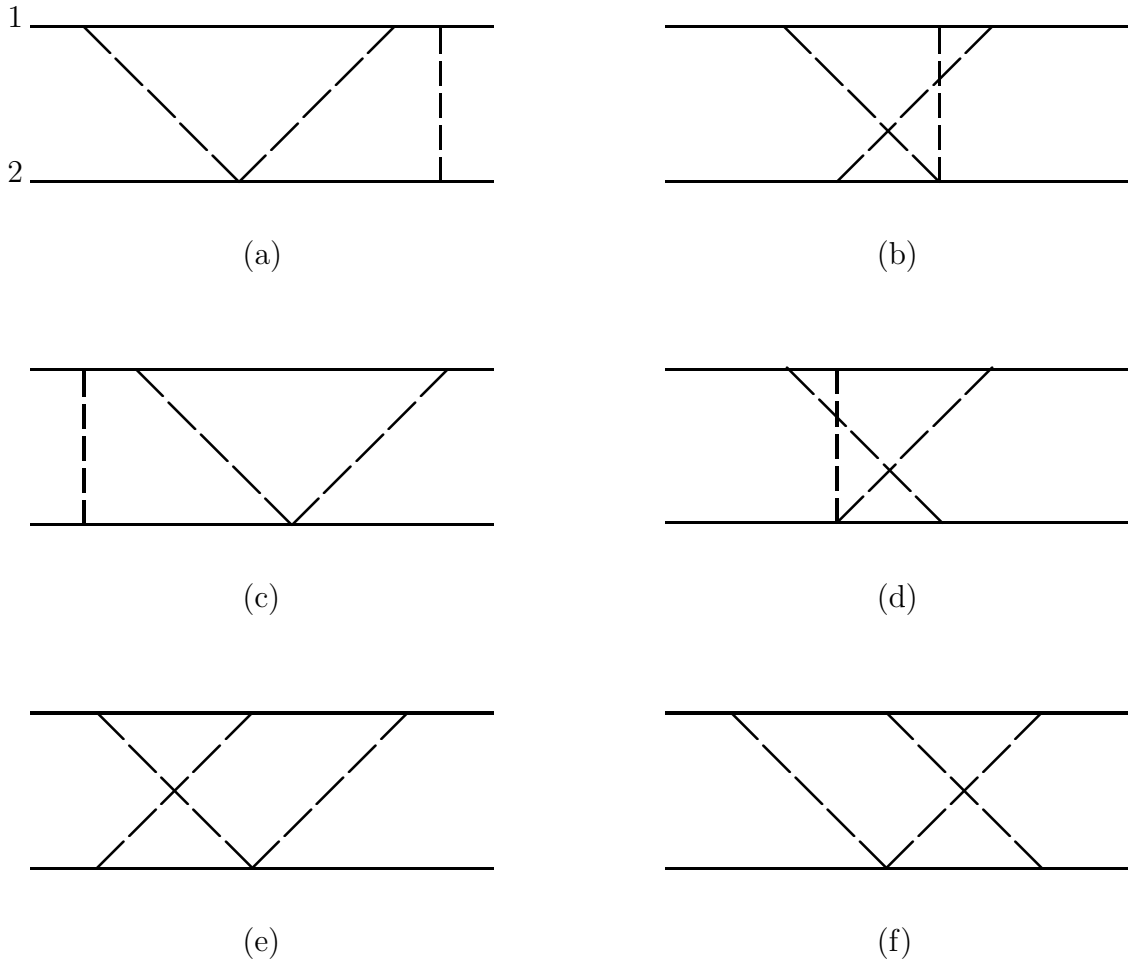
Fig. 6. Examples of mixtures of elementary diagrams in the bosonic case.
\vspace{0.3 cm}

\noindent
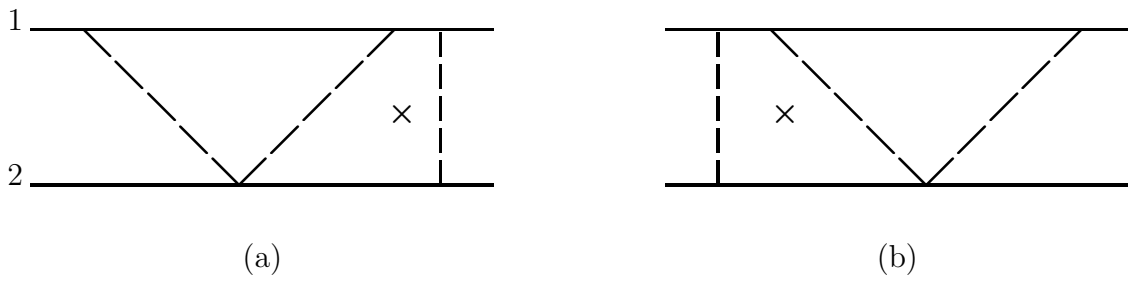
Fig. 7. ``Constraint diagrams'' associated with diagrams of Fig. 6.
\vspace{0.3 cm}

\noindent
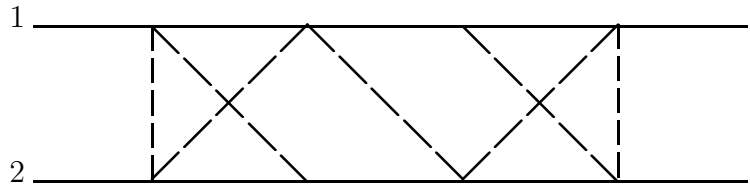
Fig. 8. Typical diagram with an odd number of exchanged scalar photons
contributing to the potential in the bosonic case.
\vspace{0.3 cm}

\noindent
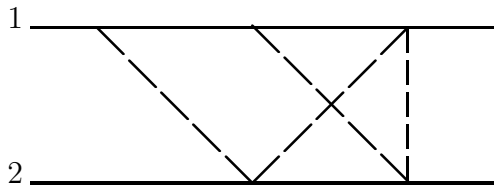
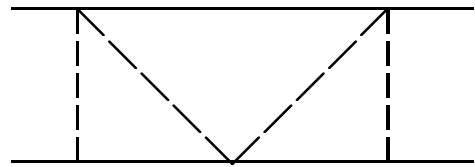
Fig. 9. Typical diagrams with an even number of exchanged scalar
photons contributing to the potential in the bosonic case.

\par

\newpage

%
%
%
%

\input FEYNMAN


\indent

\indent

\vspace{2 cm}

\begin{picture}(35000,50000)
\THICKLINES


\drawline\fermion[\E\REG](14000,48000)[5000]
\drawarrow[\LDIR\ATBASE](\pmidx,\pmidy)
\global\advance\pfrontx by -1000
\put(\pfrontx,\pfronty){$p_1$}

\drawline\photon[\S\CURLY](\fermionbackx,\fermionbacky)[5]

\drawline\fermion[\E\REG](\fermionbackx,\fermionbacky)[5000]
\drawarrow[\LDIR\ATBASE](\pmidx,\pmidy)
\global\advance\pbackx by 700
\put(\pbackx,\pbacky){$p_1'$}

\drawline\fermion[\W\REG](\photonbackx,\photonbacky)[5000]
\drawarrow[\W\ATTIP](\pmidx,\pmidy)
\global\advance\pbackx by -2000
\put(\pbackx,\pbacky){$-p_2$}

\drawline\fermion[\E\REG](\photonbackx,\photonbacky)[5000]
\drawarrow[\W\ATTIP](\pmidx,\pmidy)
\global\advance\pbackx by 200
\put(\pbackx,\pbacky){$-p_2'$}

\put(18000,37000){Fig. 1}



\drawline\fermion[\E\REG](5000,28000)[3500]
\drawarrow[\LDIR\ATBASE](\pmidx,\pmidy)
\global\advance\pfrontx by -1000
\put(\pfrontx,\pfronty){$p_1$}

\drawline\photon[\S\CURLY](\fermionbackx,\fermionbacky)[5]

\drawline\fermion[\E\REG](\fermionbackx,\fermionbacky)[3500]

\drawline\fermion[\W\REG](\photonbackx,\photonbacky)[3500]
\drawarrow[\W\ATTIP](\pmidx,\pmidy)
\global\advance\pbackx by -2000
\put(\pbackx,\pbacky){${-p_2}$}

\drawline\fermion[\E\REG](\photonbackx,\photonbacky)[3500]

\drawline\photon[\N\CURLY](\fermionbackx,\fermionbacky)[5]

\drawline\fermion[\E\REG](\fermionbackx,\fermionbacky)[3500]
\drawarrow[\W\ATTIP](\pmidx,\pmidy)
\global\advance\pbackx by 200
\put(\pbackx,\pbacky){${-p_2'}$}

\global\advance\photonbacky by 30
\drawline\fermion[\E\REG](\photonbackx,\photonbacky)[3500]
\drawarrow[\E\ATTIP](\pmidx,\pmidy)
\global\advance\pbackx by 700
\put(\pbackx,\pbacky){$p_1'$}

\put(19000,25000){\mbox{\boldmath $+$}}

\drawline\fermion[\E\REG](23000,28000)[3500]
\drawarrow[\LDIR\ATBASE](\pmidx,\pmidy)
\global\advance\pfrontx by -900
\put(\pfrontx,\pfronty){$p_1$}

\drawline\photon[\SE\FLIPPEDFLAT](\fermionbackx,\fermionbacky)[9]

\drawline\fermion[\E\REG](\fermionbackx,\fermionbacky)[5500]

\drawline\fermion[\E\REG](\photonbackx,\photonbacky)[3500]
\drawarrow[\W\ATTIP](\pmidx,\pmidy)
\global\advance\pbackx by 200
\put(\pbackx,\pbacky){${-p_2'}$}

\drawline\fermion[\W\REG](\photonbackx,\photonbacky)[5500]

\drawline\photon[\NE\FLIPPEDFLAT](\fermionbackx,\fermionbacky)[9]

\drawline\fermion[\W\REG](\photonfrontx,\photonfronty)[3500]
\drawarrow[\W\ATTIP](\pmidx,\pmidy)
\global\advance\pbackx by -2000
\put(\pbackx,\pbacky){$-p_2$}

\drawline\fermion[\E\REG](\photonbackx,\photonbacky)[3500]
\drawarrow[\E\ATTIP](\pmidx,\pmidy)
\global\advance\pbackx by 700
\put(\pbackx,\pbacky){$p_1'$}


\put(10000,13000){\mbox{\boldmath $+$}}

\put(19000,13000){\mbox{\boldmath $\times$}}
\drawline\fermion[\E\REG](14000,16000)[3500]
\drawarrow[\LDIR\ATBASE](\pmidx,\pmidy)
\global\advance\pfrontx by -900
\put(\pfrontx,\pfronty){$p_1$}

\drawline\photon[\S\CURLY](\fermionbackx,\fermionbacky)[5]

\drawline\fermion[\E\REG](\fermionbackx,\fermionbacky)[3500]

\drawline\fermion[\W\REG](\photonbackx,\photonbacky)[3500]
\drawarrow[\W\ATTIP](\pmidx,\pmidy)
\global\advance\pbackx by -2000
\put(\pbackx,\pbacky){${-p_2}$}

\drawline\fermion[\E\REG](\photonbackx,\photonbacky)[3500]

\drawline\photon[\N\CURLY](\fermionbackx,\fermionbacky)[5]

\drawline\fermion[\E\REG](\fermionbackx,\fermionbacky)[3500]
\drawarrow[\W\ATTIP](\pmidx,\pmidy)
\global\advance\pbackx by 200
\put(\pbackx,\pbacky){${-p_2'}$}

\global\advance\photonbacky by 30
\drawline\fermion[\E\REG](\photonbackx,\photonbacky)[3500]
\drawarrow[\E\ATTIP](\pmidx,\pmidy)
\global\advance\pbackx by 700
\put(\pbackx,\pbacky){$p_1'$}

\put(18000,5000){Fig. 2}

\end{picture}

\newpage


\indent

\indent

\vspace{1.75 cm}

\begin{picture}(35000,55000) 
\THICKLINES

                         
\drawline\fermion[\E\REG](1000,53000)[16500]   
\global\advance\pfrontx by -900
\put(\pfrontx,\pfronty){1}

\global\gaplength=350
\drawline\scalar[\SE\REG](3500,\fermionbacky)[5]

\drawline\fermion[\E\REG](1000,\scalarbacky)[16500]
\global\advance\pfrontx by -900
\put(\pfrontx,\pfronty){2}

\global\gaplength=350
\drawline\scalar[\NE\REG](\scalarbackx,\scalarbacky)[5]


\drawline\fermion[\E\REG](24000,53000)[16500]
\global\advance\pfrontx by -900
\put(\pfrontx,\pfronty){1}
\global\advance\pfrontx by 900
                      
\global\gaplength=350
\drawline\scalar[\SW\REG](\pmidx,\pmidy)[5]

\drawline\fermion[\E\REG](24000,\scalarbacky)[16500]
\global\advance\pfrontx by -900
\put(\pfrontx,\pfronty){2}

\global\gaplength=350
\drawline\scalar[\SE\REG](\scalarfrontx,\scalarfronty)[5]

\put(18000,42000){Fig. 3}       
                                                                                


\drawline\fermion[\E\REG](7000,33000)[28000]
\global\advance\pfrontx by -900
\put(\pfrontx,\pfronty){1}

\drawline\photon[\S\CURLY](10500,33090)[5]
\global\advance\photonfrontx by -1500
\global\advance\photonbackx by -1900
\global\advance\photonbacky by 50

\global\advance\photonbacky by 20
\drawline\fermion[\E\REG](7000,\photonbacky)[28000]
\global\advance\pfrontx by -900
\put(\pfrontx,\pfronty){2}

\drawline\photon[\S\CURLY](14000,33090)[5]
\global\advance\photonfrontx by -1500
\global\advance\photonbackx by -1500
\global\advance\photonbacky by 50

\drawline\photon[\SE\FLIPPEDFLAT](21000,33000)[9]
\global\advance\photonfrontx by -2500
\global\advance\photonbackx by -1500
                                                        
\drawline\photon[\SW\FLAT](23700,33000)[9]
\global\advance\photonfrontx by -900
\global\advance\photonbackx by -2300
                                            
\drawline\photon[\SE\FLIPPEDFLAT](26000,33000)[9]
\global\advance\photonfrontx by -600
\global\advance\photonbackx by -3000
\global\advance\photonbackx by 4500

\drawline\photon[\SW\FLAT](28700,33000)[9]
\global\advance\photonfrontx by -1100
\global\advance\photonfrontx by 4700
\global\advance\photonbackx by -1900
\put(18000,23000){Fig. 4}



\put(16000,8000){\mbox{\boldmath $\times$}}
\put(24000,8000){\mbox{\boldmath $\times$}}
\drawline\fermion[\E\REG](7000,11000)[3500]
\global\advance\pfrontx by -900
\put(\pfrontx,\pfronty){1}

\global\advance\fermionbacky by 110
\drawline\photon[\S\CURLY](\fermionbackx,\fermionbacky)[5]
\global\advance\photonbacky by 90

\global\advance\fermionbacky by -110
\drawline\fermion[\E\REG](\fermionbackx,\fermionbacky)[3500]

\drawline\fermion[\W\REG](\photonbackx,\photonbacky)[3500]
\global\advance\pbackx by -900
\put(\pbackx,\pbacky){2}

\drawline\fermion[\E\REG](\photonbackx,\photonbacky)[3500]

\global\advance\fermionbacky by -110
\drawline\photon[\N\FLIPPEDCURLY](\fermionbackx,\fermionbacky)[5]
\global\advance\fermionbacky by 110

\drawline\fermion[\E\REG](\fermionbackx,\fermionbacky)[3500]

\drawline\fermion[\E\REG](\photonbackx,\photonbacky)[3500]

\drawline\photon[\SE\FLIPPEDFLAT](\fermionbackx,\fermionbacky)[9]

\drawline\fermion[\E\REG](\fermionbackx,\fermionbacky)[5500]

\drawline\fermion[\E\REG](\photonbackx,\photonbacky)[3500]

\drawline\fermion[\W\REG](\photonbackx,\photonbacky)[5500]

\drawline\photon[\NE\FLIPPEDFLAT](\fermionbackx,\fermionbacky)[9]

\drawline\fermion[\E\REG](\photonbackx,\photonbacky)[3500]

\drawline\photon[\SE\FLIPPEDFLAT](\fermionbackx,\fermionbacky)[9]

\drawline\fermion[\E\REG](\fermionbackx,\fermionbacky)[5500]

\drawline\fermion[\E\REG](\photonbackx,\photonbacky)[3500]

\drawline\fermion[\W\REG](\photonbackx,\photonbacky)[5500]

\drawline\photon[\NE\FLIPPEDFLAT](\fermionbackx,\fermionbacky)[9]

\drawline\fermion[\E\REG](\photonbackx,\photonbacky)[3500]

\put(18000,1000){Fig. 5}

\end{picture}
\newpage                       



\indent

\indent

\vspace{1.0 cm}

\begin{picture}(35000,57000) 
\THICKLINES


\put(9000,48000){(a)}       
\drawline\fermion[\E\REG](1000,57000)[17500]   
\global\advance\pfrontx by -900
\put(\pfrontx,\pfronty){1}

\global\gaplength=310
\drawline\scalar[\SE\REG](3000,\fermionbacky)[5]

\drawline\fermion[\E\REG](1000,\scalarbacky)[17500]
\global\advance\pfrontx by -900
\put(\pfrontx,\pfronty){2}

\global\gaplength=310
\drawline\scalar[\NE\REG](\scalarbackx,\scalarbacky)[5]

\global\seglength=910
\global\gaplength=310
\drawline\scalar[\N\REG](16500,\fermionbacky)[5]


\put(33000,48000){(b)}
\drawline\fermion[\E\REG](25000,57000)[17500]

\global\gaplength=310
\drawline\scalar[\SE\REG](29500,\fermionbacky)[5]

\drawline\fermion[\E\REG](25000,\scalarbacky)[17500]

\global\seglength=910
\global\gaplength=310
\drawline\scalar[\N\REG](\scalarbackx,\scalarbacky)[5]

\global\gaplength=310
\drawline\scalar[\NE\REG](31500,\fermionbacky)[5]


\put(9000,35000){(c)}
\drawline\fermion[\E\REG](1000,44000)[17500]

\global\gaplength=310
\drawline\scalar[\SE\REG](5000,\fermionbacky)[5]
                            
\drawline\fermion[\E\REG](1000,\scalarbacky)[17500]

\global\gaplength=310
\drawline\scalar[\NE\REG](\scalarbackx,\scalarbacky)[5]

\global\seglength=910
\global\gaplength=310
\drawline\scalar[\N\REG](3000,\fermionbacky)[5]


\put(33000,35000){(d)}
\drawline\fermion[\E\REG](25000,44000)[17500]

\global\seglength=910
\global\gaplength=310
\drawline\scalar[\S\REG](31500,\fermionbacky)[5]

\drawline\fermion[\E\REG](25000,\scalarbacky)[17500]

\global\gaplength=310
\drawline\scalar[\NE\REG](\scalarbackx,\scalarbacky)[5]

\global\gaplength=310
\drawline\scalar[\NW\REG](35500,\fermionbacky)[5]


\put(9000,22000){(e)}
\drawline\fermion[\E\REG](1000,31000)[17500]

\global\gaplength=310
\drawline\scalar[\SE\REG](3500,\fermionbacky)[5]

\drawline\fermion[\E\REG](1000,\scalarbacky)[17500]

\global\gaplength=310
\drawline\scalar[\NE\REG](\scalarbackx,\scalarbacky)[5]

\global\gaplength=310
\drawline\scalar[\NE\REG](3500,\fermionbacky)[5]


\put(33000,22000){(f)}
\drawline\fermion[\E\REG](25000,31000)[17500]

\global\gaplength=310
\drawline\scalar[\SE\REG](27500,\fermionbacky)[5]

\drawline\fermion[\E\REG](25000,\scalarbacky)[17500]

\global\gaplength=310
\drawline\scalar[\NE\REG](\scalarbackx,\scalarbacky)[5]
                   
\global\gaplength=310
\drawline\scalar[\NW\REG](\pbackx,\fermionbacky)[5]

\put(19000,19000){Fig. 6}



\put(9000,3000){(a)}
\put(14500,8500){\mbox{\boldmath $\times$}}
\drawline\fermion[\E\REG](1000,12000)[17500]   
\global\advance\pfrontx by -900
\put(\pfrontx,\pfronty){1}

\global\gaplength=310
\drawline\scalar[\SE\REG](3000,\fermionbacky)[5]

\drawline\fermion[\E\REG](1000,\scalarbacky)[17500]
\global\advance\pfrontx by -900
\put(\pfrontx,\pfronty){2}

\global\gaplength=310
\drawline\scalar[\NE\REG](\scalarbackx,\scalarbacky)[5]

\global\seglength=910
\global\gaplength=310
\drawline\scalar[\N\REG](16500,\fermionbacky)[5]


\put(33000,3000){(b)}
\put(29000,8500){\mbox{\boldmath $\times$}}
\drawline\fermion[\E\REG](25000,12000)[17500]

\global\gaplength=310
\drawline\scalar[\SE\REG](29000,\fermionbacky)[5]
                            
\drawline\fermion[\E\REG](25000,\scalarbacky)[17500]

\global\gaplength=310
\drawline\scalar[\NE\REG](\scalarbackx,\scalarbacky)[5]

\global\seglength=910
\global\gaplength=310
\drawline\scalar[\N\REG](27000,\fermionbacky)[5]

\put(19000,1){Fig. 7}

\end{picture}

\newpage


\indent

\indent

\vspace{2 cm}

\begin{picture}(35000,50000) 
\THICKLINES


\drawline\fermion[\E\REG](6500,48000)[27000]
\global\advance\pfrontx by -900
\put(\pfrontx,\pfronty){1}

\global\gaplength=310
\drawline\scalar[\SE\REG](11000,\fermionbacky)[5]

\drawline\fermion[\E\REG](6500,\scalarbacky)[27000]
\global\advance\pfrontx by -900                
\put(\pfrontx,\pfronty){2}

\global\seglength=910
\global\gaplength=310
\drawline\scalar[\S\REG](\scalarfrontx,\scalarfronty)[5] 

\global\gaplength=310
\drawline\scalar[\NE\REG](\pbackx,\pbacky)[5]

\global\gaplength=310
\drawline\scalar[\SE\REG](\pbackx,\pbacky)[5]

\global\gaplength=310
\drawline\scalar[\NE\REG](\pbackx,\pbacky)[5]

\global\seglength=910
\global\gaplength=310
\drawline\scalar[\S\REG](\pbackx,\pbacky)[5] 

\global\gaplength=310
\drawline\scalar[\NW\REG](\pbackx,\fermionbacky)[5] 

\put(19000,36000){Fig. 8}



\put(9000,18000){(a)}
\drawline\fermion[\E\REG](1000,27000)[17500]
\global\advance\pfrontx by -900
\put(\pfrontx,\pfronty){1}

\global\gaplength=310
\drawline\scalar[\SE\REG](3500,\fermionbacky)[5]

\drawline\fermion[\E\REG](1000,\scalarbacky)[17500]
\global\advance\pfrontx by -900
\put(\pfrontx,\pfronty){2}

\global\gaplength=310
\drawline\scalar[\NE\REG](\scalarbackx,\scalarbacky)[5]

\global\seglength=910
\global\gaplength=310
\drawline\scalar[\S\REG](\pbackx,\pbacky)[5] 

\global\gaplength=310
\drawline\scalar[\NW\REG](\pbackx,\pbacky)[5] 


\put(33000,18000){(b)}
\drawline\fermion[\E\REG](25000,27000)[17500]

\global\seglength=910
\global\gaplength=310
\drawline\scalar[\S\REG](27500,\fermionbacky)[5]

\drawline\fermion[\E\REG](25000,\scalarbacky)[17500]

\global\gaplength=310
\drawline\scalar[\SE\REG](\scalarfrontx,\scalarfronty)[5]
                   
\global\gaplength=310
\drawline\scalar[\NE\REG](\pbackx,\pbacky)[5]

\global\seglength=910
\global\gaplength=310
\drawline\scalar[\S\REG](\pbackx,\pbacky)[5]

\put(19000,14000){Fig. 9}

\end{picture}


\end{document}